\documentclass[sigconf]{acmart}
\usepackage{listings}

\lstdefinestyle{promptblock}{
  basicstyle=\ttfamily\scriptsize,
  breaklines=true,
  breakatwhitespace=false,
  columns=fullflexible,
  keepspaces=true,
  showstringspaces=false,
  tabsize=2,
  literate=
    {×}{{$\times$}}1
    {→}{{$\rightarrow$}}1
    {←}{{$\leftarrow$}}1
    {↑}{{$\uparrow$}}1
    {↓}{{$\downarrow$}}1
    {≤}{{$\leq$}}1
    {≥}{{$\geq$}}1
    {∈}{{$\in$}}1
    {Δ}{{$\Delta$}}1
    {’}{{'}}1
    {“}{{``}}1
    {”}{{''}}1
    {–}{{--}}1
    {—}{{---}}1
}

\newcommand{\rebadd}[1]{#1}
\newcommand{\rebdel}[1]{}
\newcommand{\rebrep}[2]{#2}

\AtBeginDocument{%
  }

\copyrightyear{2026}
\acmYear{2026}
\setcopyright{cc}
\setcctype{by}
\acmConference[DIS '26]{Designing Interactive Systems Conference}{June 13--17, 2026}{Singapore, Singapore}
\acmBooktitle{Designing Interactive Systems Conference (DIS '26), June 13--17, 2026, Singapore, Singapore}
\acmDOI{10.1145/3800645.3813071}
\acmISBN{979-8-4007-2563-0/2026/06}

\begin{document}

\title{RoboBlockly Studio: Conversational Block Programming with Embodied Robot Feedback for Computational Thinking}

\author{Leyi Li}
\authornote{These authors contributed equally.}
\orcid{0009-0004-7171-3371}
\affiliation{%
  \institution{Xi'an Jiaotong-Liverpool University}
  \city{Suzhou}
  \state{Jiangsu}
  \country{China}
}
\email{Leyi.Li23@student.xjtlu.edu.cn}

\author{Chenyu Du}
\orcid{0009-0000-5192-5167}
\authornotemark[1]
\affiliation{%
  \institution{Xi'an Jiaotong-Liverpool University}
  \city{Suzhou}
  \state{Jiangsu}
  \country{China}
}
\email{Chenyu.Du23@student.xjtlu.edu.cn}

\author{Jiafei Sun}
\orcid{0009-0007-5252-3419}
\affiliation{%
  \institution{Xi'an Jiaotong-Liverpool University}
  \city{Suzhou}
  \state{Jiangsu}
  \country{China}
}
\email{Jiafei.Sun23@student.xjtlu.edu.cn}

\author{Erick Purwanto}
\orcid{0000-0001-6497-6721}

\authornote{Corresponding authors.} 

\affiliation{%
  \institution{Xi'an Jiaotong-Liverpool University}
  \city{Suzhou}
  \state{Jiangsu}
  \country{China}
}
\email{Erick.Purwanto@xjtlu.edu.cn}

\author{Qing Zhang}
\orcid{0000-0001-7296-2696}
\authornotemark[2]
\affiliation{%
  \institution{Xi'an Jiaotong-Liverpool University}
  \city{Suzhou}
  \state{Jiangsu}
  \country{China}
}
\email{Qing.Zhang02@xjtlu.edu.cn}

\renewcommand{\shortauthors}{Li et al.}


\begin{abstract}
Computational thinking (CT) is increasingly promoted as a core literacy, yet learners and teachers face challenges in connecting abstract program logic to meaningful outcomes. We design and evaluate RoboBlockly Studio, an integrated interactive system that combines block-based programming, a conversational AI teaching agent, and embodied robot execution. RoboBlockly Studio creates a tight iterative loop of authoring, running, observing, and revising. Informed by interviews with five programming teachers, the system was designed to support four goals: (1) preserving learner agency in computational thinking, (2) making program behavior transparent and interpretable, (3) grounding programming in embodied, classroom-aligned tasks, and (4) scaffolding reflection through pedagogically grounded AI dialogue. We deployed RoboBlockly Studio with 32 high school students, observing how robot and AI feedback influenced students’ interactions with code, reflections on problem-solving strategies, and understanding of CT concepts. We discuss design insights and implications for creating interactive, embodied learning environments that integrate AI and robotics to support CT learning in computing education.
\end{abstract}






\maketitle

\section{Introduction}

Computational Thinking (CT) is increasingly recognized as a core skill in K--12 education and a foundational objective in computing curricula \cite{weintrop2016defining,dong2019prada,kafai2022revaluation}, commonly described as a thinking process in which problems and their solutions are formulated so that the solutions can be effectively carried out by an information-processing agent \cite{wing2010ct}. CT comprises multiple facets; across researchers, commonly cited components include decomposition, abstraction, algorithms (algorithmic design), and debugging \cite{SHUTE2017142}. However, classroom instruction in CT is often limited to theoretical exercises or paper-based activities, offering few opportunities for learners to apply these skills to authentic problems or to benefit from embodied feedback \cite{liu2024bringing, chen2024pbl, huang2021unplugged}. This situation is also exacerbated by the fact that many teachers lack sufficient training or experience in teaching CT, making it difficult to design and facilitate engaging, hands-on learning experiences \cite{lane2023teacher,liu2024bringing,tagare2024factors}. Even when educational robots are incorporated into STEM education, their impact on students’ CT has been found to be limited and statistically non-significant \cite{ouyang2024effects}. \rebadd{In educational robotics activities, novices often struggle to move beyond trial-and-error programming toward understanding the problem and formulating an appropriate solution \cite{chevalier2020fostering}. For example, a student may assemble a program for a navigation task yet remain unsure why the robot fails to behave as intended \cite{chevalier2020fostering}. Teachers, meanwhile, may find it difficult to assess students' learning and adapt instructional support accordingly during such activities \cite{chevalier2021teachers}. Together, these conditions motivate the need for learning environments that help learners interpret program behavior more effectively while also providing accessible guidance and feedback \cite{chevalier2022role}.}

Recent advances in Human--Computer Interaction have explored conversational agents, visual programming environments, and embodied robotics as promising approaches for supporting CT learning \cite{dietz2023visual,williams2024doodlebot,chen2024chatscratch}. Prior work demonstrates that embodied interaction can make abstract concepts tangible \cite{sung2017introducing,williams2024doodlebot}, and AI-based feedback can scaffold problem-solving and debugging processes \cite{rowe2018labeling}. \rebdel{Yet, few systems integrate these strands into a unified environment that delivers structured, CT-aligned tasks, personalized guidance, and immediate robot execution.} \rebadd{However, when these supports are used separately, important limitations remain. Block-based programming environments can lower barriers to entry for novices by reducing syntax-related difficulty and supporting early engagement in programming \cite{weintrop2017comparing,kazemitabaar2022codestruct}, but this line of work primarily focuses on novice access and transition, leaving open how learners interpret program behavior in embodied settings. Embodied systems such as educational robots can make computational activities more concrete and engaging, yet prior work shows that effective CT learning depends on feedback and guidance in addition to the robot itself \cite{montuori2024cognitive,chevalier2022role}. Conversational AI can provide explanations, hints, and debugging support in programming education \cite{groothuijsen2024ai,yang2024enhancing,Wu2025TraceMate}, but existing studies mainly examine code-focused assistance, leaving open how such support should be integrated with embodied task execution. As a result, learners may still experience programming, execution, and explanation as fragmented steps rather than as an integrated process of understanding and refinement.}

To address this gap, we reviewed prior literature and conducted semi-structured interviews with five programming teachers, and developed RoboBlockly Studio, an AI-assisted visual programming platform that integrates three core components: 1) a block-based editor with a structured bank of tasks aligned to CT practices, 2) immediate execution on a physical robot, making program outcomes directly observable and actionable, and 3) a conversational teaching agent that provides adaptive, step-by-step hints and explanations. Together, these components form a closed learning loop linking task intent, program structure, and embodied robot execution feedback, enabling learners to experiment, reflect, and iteratively improve their solutions.
\rebadd{This framing also reflects our emphasis on both \textit{agentic CT} and \textit{embodied CT}. We use \textit{agentic CT} to refer to forms of computational thinking in which learners retain control over planning, revision, and verification decisions, and \textit{embodied CT} to refer to CT learning supported by observable physical robot execution that externalizes program behavior beyond the screen.}

This paper makes three contributions. Firstly, we introduce RoboBlockly Studio, a unified HCI learning environment that couples a block-based editor, a conversational teaching agent, and real-robot execution, forming a closed loop between task intent, program structure, and embodied outcomes. \rebrep{Secondly, in a counterbalanced within-subject study with high school novices (N=32), we show measurable CT learning gains and reduced workload—particularly in temporal demand, effort, and frustration—alongside good usability and positive user experience.}{Secondly, in a counterbalanced within-subject study with high school novices ($N=32$), we found overall CT learning gains across both the AI-only and AI+Robot conditions, and observed that the AI+Robot condition was associated with lower perceived workload on several NASA--TLX dimensions, alongside good usability and positive user experience.} Finally, we discuss design guidelines for AI-supported, robot-mediated CT instruction, highlighting how embodiment provides perceptual feedback that sharpens debugging and supports learner agency while conversational hints scaffold reflection.

\begin{figure}[t]
  \centering
  \includegraphics[width=\linewidth]{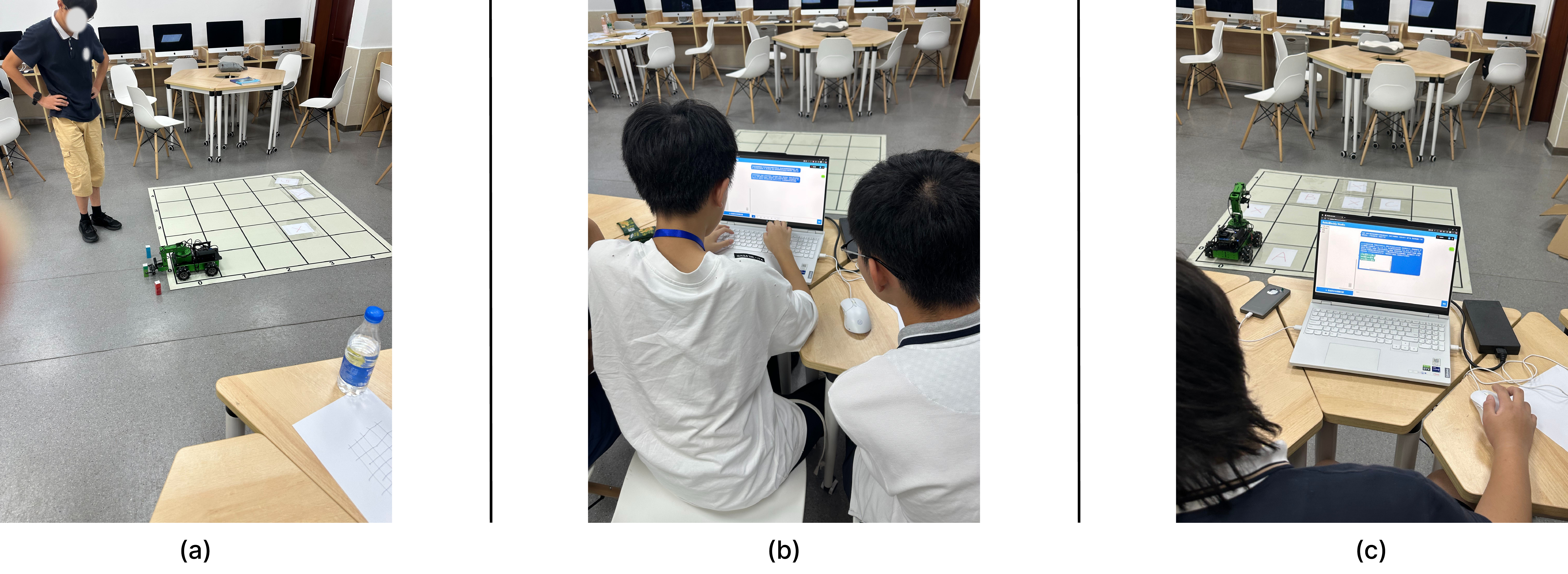}
  \caption{(a) observing the robot in operation, (b) engaging in problem-solving, and (c) examining the map to plan the path.}
  \Description{The figure contains three photos labeled (a)–(c) showing a classroom robotics programming activity. (a) A small wheeled robot sits at the edge of a taped, 5$\times$5 grid mat on the floor; a student stands nearby while desks and computers are visible in the background. (b) Two students sit side-by-side at a desk, looking at a computer screen and using a mouse/keyboard as they work on RoboBlockly Studio; the grid mat is visible on the floor behind them. (c) A laptop on a desk displays the RoboBlockly Studio interface while the robot and grid mat are visible nearby, indicating students program on the computer and then test code on the physical robot.}
  \label{fig:picture}
\end{figure}

\begin{figure*}[t]
  \centering
  \includegraphics[width=0.95\textwidth]{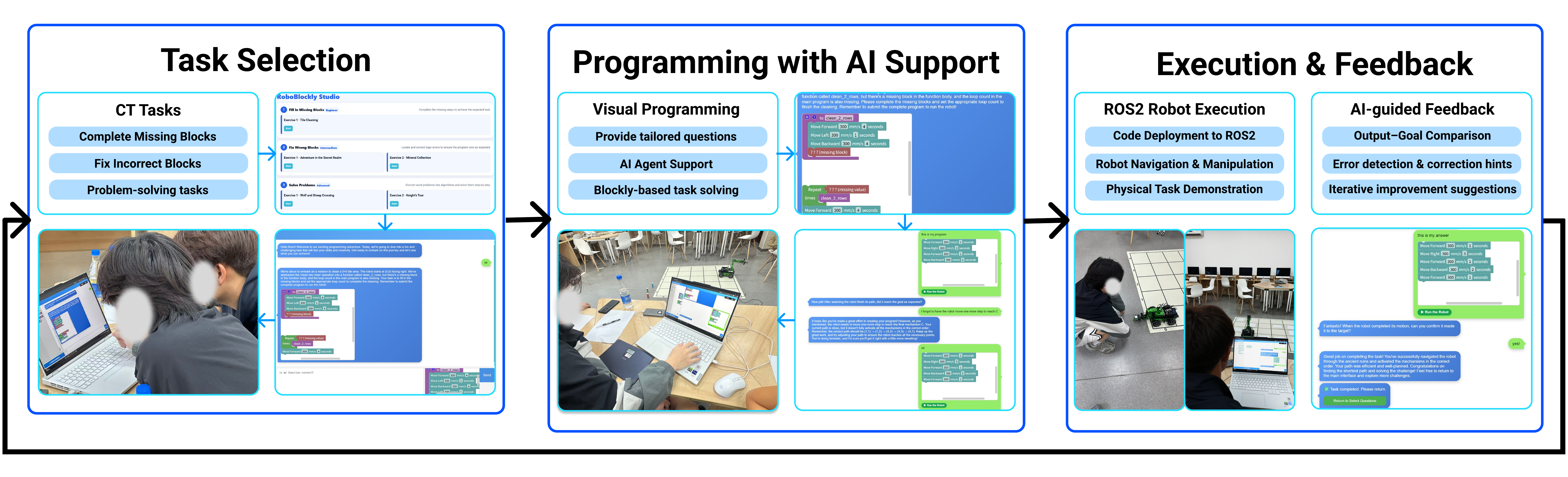}
  \caption{An overview of RoboBlockly Studio. The framework integrates block-based programming, conversational AI, and embodied robot execution, allowing students to design, test, and refine programs with immediate physical feedback.}
  \Description{Three-panel workflow diagram of RoboBlockly: Task Selection (students choose computational-thinking tasks such as completing missing blocks, fixing incorrect blocks, or problem-solving tasks), Programming with AI Support (students build block-based code with an AI agent that asks tailored questions and supports task solving), and Execution & Feedback (code is deployed to a ROS2 robot for a physical demonstration, while the AI provides output–goal comparison plus error-correction and iterative improvement suggestions).}
  \label{fig:architecture}
\end{figure*}

\section{BACKGROUND AND RELATED WORK}

\subsection{Computational Thinking in K--12 and HCI}
Computational thinking research for K--12 education has shifted from simply making programming accessible to designing environments that make learners’ problem-solving and conceptual reasoning visible. Early systems such as CTArcade engaged children in authoring and debugging rule-based strategies through iterative play~\cite{lee2012ctarcade}. Rowe et al.~\cite{rowe2018labeling} showed how gameplay logs can capture implicit CT practices for assessment. More recent work emphasizes multimodal and embodied perspectives: Zaman~\cite{zaman2023exploring} analyzed gestures in collaborative reasoning, framing CT as distributed across artifacts and representations. Williams et al.~\cite{williams2024doodlebot} introduced Doodlebot, linking hands-on experimentation with abstraction and debugging. Efforts to broaden inclusion include robotics activities for neurodiverse preschoolers~\cite{das2025cultivating} and teacher-in-the-loop assessment of Scratch projects~\cite{troiano2025ct4all}. 

A second line of work foregrounds the orchestration constraints that shape CT learning tools in real classrooms: teachers must maintain lesson tempo while diagnosing heterogeneous misconceptions and offering differentiated support in-the-moment. Prior DIS studies show that peripheral and teacher-facing awareness systems can help teachers notice where support is needed and time interventions without derailing whole-class flow (e.g., peripheral visualizations during classroom activity; differentiated-instruction probes) \cite{10.1145/3322276.3322365,10.1145/3357236.3395497}. Some data-driven classroom technologies can introduce bias and surveillance-like dynamics, making careful boundary-setting and minimally disruptive design essential \cite{10.1145/3461778.3462084,10.1145/3563657.3596079}. Complementing this, inclusive computing work reminds us that “access” is not only about lowering technical barriers, but also about supporting diverse abilities and participation structures \cite{10.1145/3173574.3174091}.

Collectively, prior work suggests CT tools should externalize execution and intermediate representations to ground discussion in shared evidence, while offering differentiated scaffolding that preserves learners' ownership. RoboBlockly Studio addresses this by supporting a rapid author–run–observe–revise loop: embodied enactment makes behavior observable, and a conversational agent delivers reflective prompts. Together, tangible traces and dialogue-based reflection support learners' strategy articulation and enable teachers to diagnose misconceptions beyond final answers.

\subsection{Conversational Agents and Block-Based Programming}
Conversational agents have been leveraged for scaffolding programming learning through natural language. Robe and Kuttal~\cite{robe2022designing} proposed PairBuddy, modeling conversational moves for pair programming. Chen et al.~\cite{chen2024chatscratch} designed ChatScratch, augmenting Scratch with planning and LLM-based support. Large-scale deployments such as CodeAid~\cite{kazemitabaar2024codeaid} show that assistants can aid understanding and debugging while preserving agency. Dialogue coupled with algorithm visualizations deepens comprehension~\cite{li2025visualcodemooc}, and simulated learner interactions train tutoring skills~\cite{pan2025tutorup}. Across this body of work, dialogue is commonly treated as a mechanism for sensemaking rather than mere answer delivery. Bernuy et al.\cite{10.1145/3643834.3661596} recommend that prompts for self-explanation can externalize intermediate thinking, including in voice-mediated settings. Hinting interfaces commonly provide progressive, multi-level support that can be tapered as learners gain competence \cite{10.1145/3613905.3650937,10.1145/3411764.3445228,10.1145/2702123.2702580}. In addition, many tutoring-style interactions can be framed as a lightweight debugging loop of interpreting failures, proposing causes, attempting repairs, and validating revisions, so conversation supports iterative hypothesis testing rather than one-off correction \cite{10.1145/3706598.3714002,10.1145/3313831.3376494,10.1145/3313831.3376857}. Together, these works establish learning-by-chatting as a promising paradigm, though interactions typically remain confined to the screen without embodied execution.

Block-based programming environments are now common in K--12, in part because their puzzle-piece constraints prevent many syntactic errors and let novices focus on program structure rather than surface form \cite{10.1145/3173574.3173643,10.1145/3025453.3025711}. From an HCI perspective, effective learning environments typically require iterative, stakeholder-informed refinement so that representations, constraints, and feedback align with learners’ evolving mental models and the realities of classroom orchestration \cite{10.1145/3544549.3573863,10.1145/3544549.3585775,10.1145/3706599.3719763}. Within this design space, prior work has diversified that block-based programming environments lower entry barriers and foster creativity. Visual StoryCoder integrates multimodal input~\cite{dietz2023visual}, while participatory co-design approaches empower students to shape block-based tools~\cite{limke2023empowering}. Expanding into robotics, Duplo proved more efficient for novices than RobotStudio Online~\cite{fronchetti2024block}. Inclusive studies show workspace-awareness cues supporting mixed-ability collaboration~\cite{rocha2025awareness}. More recently, LLMs have been used to translate natural-language prompts into editable block pipelines, accelerating prototyping and structured hinting~\cite{zhou2025instructpipe}. Despite these advances, most environments stop at block construction and simulation, lacking embodied integration or adaptive conversational support.

In essence, prior conversational supports improve planning and debugging, and block-based environments lower syntactic barriers; however, learners still struggle to connect conversational advice to concrete behavioral evidence beyond the screen.
RoboBlockly Studio integrates block previews, optional execution, and robot-grounded feedback into the dialogue loop, enabling students to validate hypotheses through embodied outcomes while retaining control over when and how AI support is applied.

\subsection{Embodied AI and Educational Robotics}
Embodied and tangible learning research argues that coupling conceptual reasoning with perceptible, sensorimotor action and timely feedback can make abstract ideas more graspable for learners \cite{chatain2023embodied,revelle2005tui,zaidi2025tangibuild}. 
In the programming domain, tangible programming systems illustrate how children can construct programs as physical artifacts and only later compile or execute them, supporting co-located collaboration and off-screen reasoning around program structure \cite{horn2006tangible,horn2007tern}. 

Work on embodied robot and educational robotics shows how robots can make abstract concepts tangible and sustain engagement. PopBots demonstrated early-childhood AI learning gains through hands-on interaction~\cite{williams2019artificial}. Cross-representation transfer was observed from tangible to block-based coding~\cite{pedersen2021educational}, while concept-learning perspectives emphasized variation and analogical reasoning~\cite{booth2022revisiting}. Interactive teaching has been shown to rebuild trust after errors~\cite{chi2024interactive}, and PhysioBots link physiological signals to robot actions for reflective inquiry~\cite{lewis2025physiobots}. This literature underscores the value of robot embodiment.

Overall, embodied robotics and tangible interaction can render abstract computational ideas experientially concrete, supporting trust repair and post-error reflection. However, many educational robot systems still lack an integrated loop that structures CT practice through progressive tasks, scaffolds reasoning through dialogue, and anchors debugging in immediate, observable physical execution. RoboBlockly Studio introduces such a unified loop, enabling learners to iteratively test, explain, and revise programs with robot-grounded evidence while preserving agency.

\begin{table*}[t]
\centering
\caption{Design goals for CT learning environments.
\newline \footnotesize{Theoretical principles (DG1--DG2) are grounded in prior literature; empirical principles (DG3--DG4) are synthesized from teacher interviews.}}
\Description{Table with three columns: Category, Principle, and Design Consideration—listing four design goals for computational thinking (CT) learning environments: DG1 (CT as Agentic Practice): treat CT as situated, executable practice supporting problem formulation, iterative refinement, and learner control; DG2 (Transparent and Scaffolded Learning): make program behavior visible/interpretable and increase task complexity to scaffold reasoning and debugging; DG3 (Embodied and Classroom-Aligned Design): align digital program constructs with embodied actions and classroom practices for seamless integration; DG4 (Pedagogically Grounded AI Mediation): provide interpretable, instructionally grounded AI feedback that preserves learner ownership rather than replacing students’ problem solving.}
\label{tab:ct-design}
\begin{tabular}{@{}p{0.10\textwidth}p{0.24\textwidth}p{0.58\textwidth}@{}}
\toprule
\textbf{Category} & \textbf{Principle} & \textbf{Design Consideration} \\
\midrule

\textbf{\textit{DG1}} &
CT as Agentic Practice &
CT learning environments should treat computational thinking as a set of situated, executable practices, supporting problem formulation, iterative refinement, and learner control over solution strategies. \\

\midrule
\textbf{\textit{DG2}}&
Transparent and Scaffolded Learning &
Systems should make program behavior visible and interpretable, while progressively increasing task complexity to scaffold reasoning, debugging, and deeper CT engagement. \\

\midrule
\textbf{\textit{DG3}} &
Embodied and Classroom-Aligned Design &
Learning environments should align digital program constructs with embodied actions and classroom practices, enabling seamless integration into teachers’ instructional rhythms. \\

\midrule
\textbf{\textit{DG4}}&
Pedagogically Grounded AI Mediation &
AI support should provide interpretable, instructionally grounded feedback that preserves learner ownership rather than replacing students’ problem-solving processes. \\

\bottomrule
\end{tabular}
\end{table*}

\section{ROBOBLOCKLY STUDIO}
To inform the development of RoboBlockly Studio, we synthesized core design goals based on both prior literature and semi-structured interviews. These design goals, summarized in Table~\ref{tab:ct-design}, guided the system's development. Through RoboBlockly Studio, students can construct programs, interact with an AI agent, and observe the physical robot’s movements to solve CT tasks. 

\subsection{Design Considerations}
We reviewed relevant literature on CT task design, tangible programming environments, and AI-supported learning, extracting insights that we refer to as “theoretical design requirements.” These requirements guided the overall direction and high-level specifications of our system, helping to define both its conceptual scope and key functional priorities.

In parallel, we conducted interviews with five computer science instructors, including three high school teachers and two university professors. \rebadd{Across interviews, four recurring concerns emerged. First, teachers reported that students often struggled to connect block sequences to downstream behavior, especially during debugging. Second, teachers emphasized the need for progressive task structures that fit short classroom sessions and heterogeneous learner pacing. Third, they described physical execution as valuable when it made failures inspectable, but only when classroom orchestration remained lightweight. Fourth, they considered AI support useful when it offered hints, explanations, and verification prompts rather than directly replacing students’ problem solving.} Focusing on CT instruction, we designed an open-ended questionnaire complemented by in-person discussions to elicit their perspectives. Details of the questionnaire are available in Appendix B. Insights from these interactions were summarized as “empirical design requirements.”

To synthesize the theoretical and empirical findings, three authors employed a cross-coding process: insights from the literature and interviews were systematically compared, clustered, and iteratively refined to identify overlapping themes and complementary considerations \cite{FeredayMuirCochrane2006}. \rebadd{These interview findings most directly sharpened DG3 and DG4, while DG1 and DG2 were additionally grounded in prior literature on CT, learner agency, and scaffolded learning.} This analysis led to the articulation of four design goals that guided the development of RoboBlockly Studio and discussed in the next subsection. Table~\ref{tab:ct-design} provides a summary of both theoretical and empirical design considerations.

\subsection{Design Goals}
\label{subsec:design-goals}

\paragraph{\textbf{DG1: Computational Thinking as Agentic Practice.}} 
We conceptualize computational thinking not merely as the execution of predefined programming steps, but as an agentic practice through which learners actively engage with problem spaces, consistent with the notion of computational thinking as a core problem-solving approach \cite{wing2010ct}. Within this perspective, students actively formulate problems, explore alternative solution strategies, construct and test their own solutions, and iteratively refine their approaches based on feedback \cite{weintrop2016defining}. Moreover, CT as agentic practice emphasizes learner autonomy, highlighting that students maintain control over program structure, decision-making, and verification processes, rather than passively following instructions or consuming system-generated answers \cite{code2020agency}.

In practice, this means that CT learning environments should provide tools that are interpretable, manipulable, and responsive, allowing learners to experiment, observe outcomes, and adjust strategies dynamically \cite{weintrop2016defining}. For example, a student might modify the parameters of a function, test the resulting program on a simulated robot, identify unexpected behaviors, and iteratively correct the logic. By supporting this cycle of experimentation and reflection, agentic CT practices foster not only technical skill in programming but also higher-order reasoning, problem-solving, and metacognitive awareness, enabling learners to take ownership of both the process and the outcomes of their computational work \cite{wing2010ct,code2020agency}.

\paragraph{\textbf{DG2: Transparent and Scaffolded Learning.}}
Computational thinking develops most effectively when learners can observe and interpret program behavior and reasoning processes \cite{resnick2009scratch}. Learning environments should therefore provide transparency by exposing program execution, intermediate states, and the consequences of learner actions, allowing students to reason about cause and effect \cite{SHUTE2017142}. At the same time, structured scaffolding should progressively increase task complexity, supporting learners in moving from basic task completion to more sophisticated reasoning, debugging, and reflective practices \cite{wang2020crescendo}. By combining transparency with scaffolded progression, CT environments empower students to understand, experiment with, and refine their computational strategies, fostering deeper engagement and learner autonomy.

\paragraph{\textbf{DG3: Embodied and Classroom-Aligned Design.}}
To foster meaningful computational thinking learning in K--12 classrooms, digital programming environments should be closely aligned with embodied actions and existing instructional practices, as highlighted by multiple teacher interviews (T1--T5). Embodied interaction allows learners to connect abstract program logic with observable, perceptible outcomes, making computational processes tangible and easier to reason about (T1--T3). At the same time, classroom-aligned design ensures that activities fit within teachers’ time constraints, lesson structures, and pedagogical goals (T4--T5). By integrating digital programming with embodied and contextual classroom practices, learning environments make CT practices both observable and discussable, supporting collaboration, reflection, and meaningful engagement in authentic educational settings. \rebadd{These interview findings motivated DG3 by indicating that learners need observable execution outcomes, but that embodied activities must also remain manageable within classroom routines.}

\paragraph{\textbf{DG4: Pedagogically Grounded AI Mediation.}}
AI support in computational thinking learning environments should be pedagogically grounded and interpretable, providing guidance that enhances rather than replaces learners’ problem-solving processes (T1--T5). According to teacher interviews, effective AI mediation scaffolds reflection, explanation, and verification while preserving learner agency (T2, T3, T4). This means that learners remain in control of decision-making, using AI feedback to inform and adjust their strategies rather than passively accepting system-generated solutions. Moreover, AI support should respect instructional boundaries and complement teachers’ roles, providing prompts or hints that learners can interrogate, accept, or reject, thereby integrating seamlessly into classroom practices and reinforcing student ownership of both process and outcomes (T1, T4, T5). \rebadd{These interview findings motivated DG4 by indicating that AI should scaffold reflection and verification while preserving learner agency, rather than directly solving tasks for students.}

\subsection{System Overview}
RoboBlockly Studio is a web-based framework that integrates block-based visual programming, conversational AI, and robot embodiment to enhance the development of computational thinking skills (see Fig.~\ref{fig:architecture}). The system features a dual-panel interface for both programming and dialogue, while the backend converts visual code into Python and transmits it over the network to the robot’s onboard computer, which manages the control of the physical platform. The robot, equipped with a Mecanum-wheeled chassis for omnidirectional mobility and a multi-joint arm for manipulation, facilitates embodied interaction scenarios that bridge the gap between abstract program logic and tangible behavior. In this way, RoboBlockly promotes essential aspects of computational thinking through iterative cycles of programming, execution, and reflection, all situated within a human–computer interaction context.

\subsection{Computational Thinking Tasks}

Before engaging in programming and dialogue, students first select a task from a structured set of computational thinking challenges. Progressively reducing instructional support through tiered scaffolding allows novice learners to build programming skills within a controlled and stepwise learning environment \cite{noordin2025computational}. The tasks are designed to scaffold core competencies such as abstraction, decomposition, debugging, and algorithmic design \cite{grover2013ct}, and are organized into three progressive levels of difficulty \textbf{\textit{(DG2)}}:

\textbf{Introductory tasks} require students to complete partially provided programs by inserting missing blocks. For example, given predefined functions and a main routine, students add the necessary blocks to ensure the robot traverses the entire grid.

\textbf{Intermediate tasks} require students to analyze programs with deliberate flaws. For example, a predesigned path for the robot is not the shortest due to logical errors and obstacles, and students must identify these flaws and correct the program to achieve the optimal solution. Similarly, in the \textit{Minefield Exploration} task, learners plan a route to collect hidden ores on a 5$\times$5 grid while managing limited energy and avoiding swamps, correcting inefficient or unsafe paths.

\textbf{Advanced tasks} present text-based problems that require students to translate natural language into executable block-based programs. For instance, in the \textit{Wolf, Sheep, and Cabbage} river crossing puzzle, students design a transport strategy that respects safety constraints, while in the \textit{Knight’s Chessboard Challenge}, they plan a sequence of L-shaped moves on a 5$\times$5 grid to cover all non-obstacle cells and reach a target. Both tasks require learners to reason about constraints, sequence actions, and achieve goal states effectively.

Across all levels, these tasks are designed to support computational thinking as an agentic practice \textbf{\textit{(DG1)}}. By engaging with problems that require students to plan, test, and iteratively refine their solutions, learners exercise autonomy over program structure, decision-making, and verification. 

\subsection{User Scenario}

\begin{figure*}[t]
  \centering
  \includegraphics[width=0.95\linewidth]{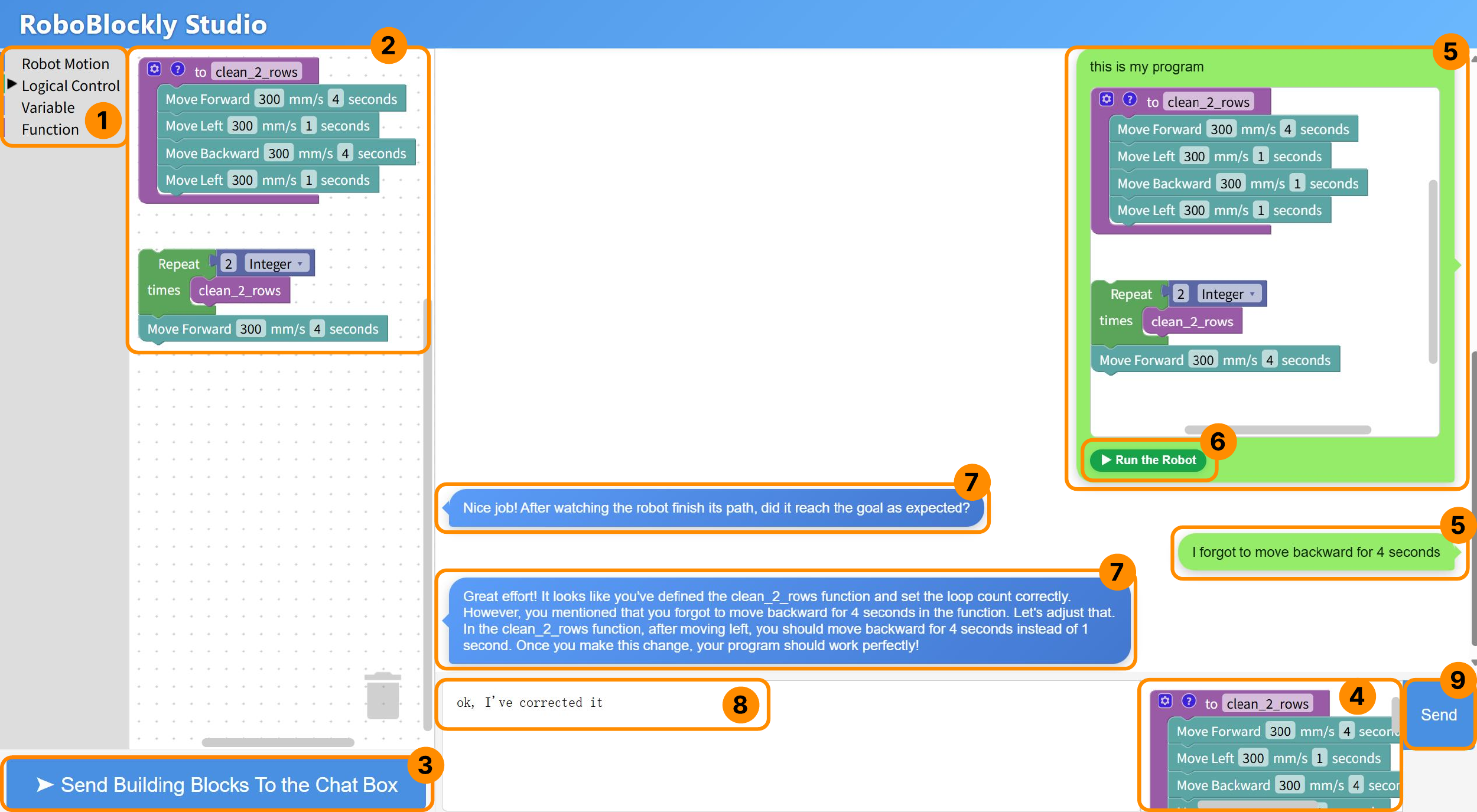}
  \caption{User Interface of RoboBlockly Studio. (1) Block Category Selection. (2) Workspace. (3) Send to Chat. (4) Block Preview. (5) Program and Messages Panel. (6) Run Robot. (7) AI Feedback. (8) Integrated Message Input. (9) Send Message.}
  \Description{Screenshot of RoboBlockly Studio showing a dual-panel interface with numbered callouts: (1) block category menu (e.g., robot motion, logic, variables, functions) and (2) a workspace where a block-based program is assembled (including a function and repeat loop). A button (3) sends the constructed blocks to the chat. On the right, a program-and-messages panel (5) displays a block preview (4) inside the conversation, a “Run the Robot” button (6) within the program bubble, and AI feedback (7) responding to the student’s message. At the bottom are an integrated message input (8) and send message button (9).}
  \label{fig:interface}
\end{figure*}

To illustrate how the system operates in practice, we present an imagined scenario involving Robby, a high school student learning computational thinking through programming. After selecting a task, Robby enters the program construction studio, a block-based environment that integrates visual coding with conversational interaction (see Fig.~\ref{fig:interface}). The left side of the interface provides categorized coding blocks and a workspace for assembling executable programs, while the right side hosts a conversational AI panel that can expand when needed. Within this panel, learners interact with the AI through both natural language and block previews, enabling them to exchange prompts, submit programs for execution on the robot, receive feedback, and preserve a record of their reasoning process. To support flexible engagement, the interface also allows dynamic resizing: clicking the chat panel background enlarges the code workspace and shrinks the chat window when greater focus on code construction is needed.

This interactive configuration emphasizes a balance between programming practice and reflective dialogue \textbf{\textit{(DG1)}}. Learners engage with core concepts such as sequencing, loops, and modular decomposition while benefiting from AI-mediated guidance, thereby linking abstract problem descriptions with embodied robot execution through iterative cycles of coding and conversation. In the following sections, we examine the key design features of RoboBlockly Studio, and present example scenarios showing how Robby leverages these features to iteratively refine his program based on feedback.

\subsubsection{AI-Guided Dialogue and Verification}

The AI dialogue module serves as an intelligent cognitive scaffold that supports learners by addressing questions, resolving uncertainties, and guiding reflective reasoning and self-assessment. Learners can engage in dialogue with the AI at any stage of program construction, either to clarify doubts or to seek confirmation. Once a program is submitted to the conversational interface, its block-based representation is displayed within the chat bubble, accompanied by a “Run the Robot” button. Learners may choose to click this button to execute the program on the robot, observe the resulting behavior, and then reflect on the outcome in response to the AI’s inquiry regarding task completion. \rebrep{The AI subsequently evaluates the correctness of the solution and provides targeted guidance \textbf{\textit{(DG4)}}.}{The AI subsequently comments on the current solution state and provides targeted guidance \textbf{\textit{(DG4)}}. In our implementation, pedagogical feedback and task checking are both handled by the LLM. To reduce unconstrained model judgment, we provide task-specific prompts that explicitly encode the relevant rules, constraints, and success criteria for each task.} Alternatively, learners may opt to submit a program without execution, using the conversational exchange with the AI to iteratively refine their code prior to running the robot \textbf{\textit{(DG2)}}. Learner agency reflects the ability to manage, direct, and monitor one’s own learning activities~\cite{code2020agency}. By allowing students to decide whether to run or merely discuss their program, the environment enhances learner agency and fosters a stronger sense of control over the problem-solving process. 

\paragraph{\textit{Example Scenario}.}  
Robby selects an introductory task involving cleaning a tiled area to protect the school environment. The problem provides a predefined function, but it contains missing blocks in its body and no specified loop count in the main program. Robby begins by reasoning through the task, completing the function and the main program by filling in the missing blocks, then submitting them to the conversational interface. Instead of immediately running the robot, he first requests AI verification. The AI highlights an error in his loop count and offers supportive feedback. After reflecting, Robby corrects the main program and resubmits his solution. Upon clicking the “Run the Robot” button, he observes the robot successfully cleaning the entire area, and the AI confirms completion and provides a button to return to the task selection interface.

\subsubsection{Real-Time Robot Execution and Feedback}

After submitting their program through the conversational interface, learners can initiate execution via a “Run the Robot” button embedded in the dialogue bubble. The block-based programs are automatically translated into Python code and transmitted to the robot through a WebSocket connection. On the robot side, a pre-initialized ROS 2 node receives the code and executes the corresponding actions on the hardware. The robot then performs the programmed actions, providing visual, kinesthetic, and spatial cues that concretize abstract concepts such as sequencing, iteration, and decomposition, enabling real-time interaction. \rebadd{This interaction refers to the immediate robot feedback provided after execution is triggered by the Run button. During program execution, the robot cannot be interrupted or controlled. Instead, learners observe the outcome, consult the AI, revise their code, and continue the interaction through subsequent runs.} \textbf{\textit{(DG3)}}.

\paragraph{\textit{Embodied Task Representation}.}
Observation of the robot’s behavior on a 5$\times$5 grid, augmented with soft PVC panels and stickers representing obstacles, targets, and special objects, as well as differently colored cubes representing manipulable objects, provides a concrete and visually intuitive representation of task dynamics. This setup allows students to form mental models of robot mechanics, object interactions, and underlying program logic.

\paragraph{\textit{Iterative Feedback and Refinement}.}
The hands-on feedback loop promotes experiential learning, as learners iteratively compare predicted outcomes with actual robot performance, identify discrepancies, and refine their programs accordingly. Situating these experiences within authentic tasks aligns with cognitive apprenticeship principles, making abstract computational thinking skills more explicit and enabling students to connect their learning to real-world applications \cite{fennell2019computational}.

\paragraph{\textit{Example Scenario}.}
Robby selects an intermediate task situated in an ancient ruin, requiring the robot to sequentially activate three mechanisms, along the shortest path while avoiding collapsed areas. A candidate block-based path is provided, which Robby identifies as suboptimal. He then constructs his own path, submits it, and immediately executes it on the robot. During execution, the robot’s actions reveal that the second step traverses a blocked area, delivering real-time, observable feedback. Utilizing this feedback, Robby detects the error, revises his program, and resubmits it. Upon re-execution, the robot successfully follows the correct path, and the AI provides confirmation of task completion.

\subsection{Implementation}
\label{subsec:implementation}

To realize the approach described above, we developed RoboBlockly Studio, a web-based learning environment implemented with Java\-Script and Node.js. The system integrates block-based visual programming, conversational AI (GPT-4o via the OpenAI API), and robot embodiment to support CT skill development. Learners author programs in the browser, where block programs are converted into executable Python and transmitted via WebSocket to the robot’s onboard Jetson Orin Nano running ROS~2 for physical execution.

We use GPT-4o to support multiple sub-tasks in the learning workflow, including (i) checking CT tasks and constraints, (ii) checking learner submissions and producing block-grounded feedback, and (iii) providing hints that scaffold planning and debugging without directly replacing learners’ reasoning. Following prior approaches that decompose complex interactions into LLM primitive operations, our prompts are organized into three parts: (1) setting an instructional role and tone (e.g., a supportive CS/robotics tutor), (2) providing explicit instructions and output constraints for the target sub-task, and (3) injecting up-to-date interaction context (e.g., task specification, available blocks, learner program state, and execution results) captured from user actions. For hint-oriented prompts, we additionally include descriptive criteria to encourage reflective reasoning (e.g., prompting learners to articulate constraints, hypothesize causes of failure, and propose minimal revisions). \rebdel{Sample prompts are provided in Appendix A.}\rebadd{Representative prompts used for correctness checking and pedagogical scaffolding are provided in Appendix A.}

\rebadd{GPT-4o was not fine-tuned for our tasks; we used it through task-specific prompt engineering with temperature set to 0. Correctness is assessed by the LLM against task-specific rules, constraints, and success criteria specified in the prompt, rather than by exact program matching, since multiple Blockly programs may validly solve the same task.}

\rebadd{The model always receives the task prompt together with a rolling window of recent textual interaction history, truncated by removing the oldest entries once it exceeds 20 messages. In addition, the current Blockly workspace is represented as XML text on the backend and incorporated into the interaction history, allowing feedback to be grounded in the learner’s evolving program state.}

\section{USER STUDY}
\label{sec:userstudy}
To evaluate the usability of RoboBlockly Studio in supporting novice learners’ computational thinking and programming experiences, we conducted a within-subject user study with 32 high school students, which received the ethical approval from authors’ institution. The study was guided by three research questions: 

\begin{itemize}
\item \textbf{RQ1:} What is the impact of embodied robot execution on learners’ programming and debugging strategies?
\item \textbf{RQ2:} How does the integration of AI guidance with visual programming influence learners’ problem-solving processes?
\item \textbf{RQ3:} In what ways can a unified environment linking task goals, program structures, and embodied feedback support the development of CT skills?
\end{itemize}

\subsection{Participants}
We recruited 32 high school students (\textit{N} = 32, age 15--18) from local schools, none of whom had prior programming experience. Participants were randomly assigned into two groups of equal size. One group first engaged with the \textit{AI--Only} condition (RoboBlockly Studio with AI support but without robot embodiment), whereas the other group began with the \textit{AI+Robot} condition (RoboBlockly Studio with both AI support and robot embodiment). This \textit{counterbalanced}, within-subject design ensured that all participants interacted with both conditions while reducing bias from sequence effects.

\subsection{Measures}
All participants completed a series of assessments combining performance, usability, and experiential measures. To evaluate computational thinking, we administered eight Bebras problems (Junior group, ages 14–16)\footnote{\url{https://bebraschallenge.org/}} \cite{dagiene2008bebras,dagiene2016s}: four in a pre-test and four in a post-test, each set including one easy (A-level), two medium (B-level), and one hard (C-level) task. \rebdel{Subjective workload was measured with the NASA Task Load Index (NASA--TLX), which captures six dimensions of workload: mental, physical, and temporal demand, as well as perceived performance, effort, and frustration \cite{cao2009nasa,hart1988nasa}.} \rebdel{Following common practice, we reverse-coded the NASA--TLX Performance subscale so that higher values consistently indicate higher workload across all six dimensions (i.e., $Performance_{rev} = 20 - Performance_{raw}$).} \rebadd{Subjective workload was measured with the NASA Task Load Index (NASA-TLX), which captures six dimensions of workload: mental, physical, and temporal demand, as well as perceived performance, effort, and frustration \cite{cao2009nasa,hart1988nasa}. In the standard NASA--TLX format, the Performance subscale is anchored from \textit{Perfect} to \textit{Failure}, so higher raw scores indicate poorer self-rated performance. We therefore report the Performance subscale without reverse-coding.} System usability was assessed using the System Usability Scale (SUS) \cite{lewis2018system,bangor2008empirical}, and overall user experience was measured with the User Experience Questionnaire (UEQ), covering perceptions of attractiveness, perspicuity, efficiency, dependability, stimulation, and novelty \cite{laugwitz2008ueq,hinderks2017design}. Finally, we conducted semi-structured interviews and collected video recordings to capture students’ experiences, reasoning strategies, and reflections, which were analyzed thematically \cite{adeoye2021research}.

\subsection{Materials}
\rebrep{The study was conducted in a computer lab classroom.}{The study was conducted in a classroom setting.} Each participant used an individual desktop computer to access RoboBlockly Studio, enabling the AI--Only and AI+Robot groups to work in parallel on their assigned condition during each session. \rebadd{Both groups were in the same classroom. To prevent cross-condition exposure, the single robot was placed at a physically separated station so that participants could not observe executions outside their assigned condition.} RoboBlockly Studio provided three types of block-based tasks, along with an AI teaching agent that offered on-demand hints, stepwise guidance, and reflective debugging support.

For the embodied robot condition, participants interacted with a ROS2-controlled robot, consisting of a mobile base, robotic arm, and gripper, on a custom-designed scenario mat. The mat incorporated physical elements such as obstacles, switches, goal regions, and manipulable objects, allowing students to engage with tasks in a tangible, contextualized environment.

\subsection{Procedure}
The study followed a structured five-stage protocol conducted in the classroom. At the outset, participants were welcomed, introduced to the study, and completed informed consent procedures; written consent was also obtained from their parents or legal guardians as well as approval from their teachers. They then completed a 15-minute Bebras pre-test consisting of four problems, designed to assess baseline computational thinking skills.

Next, every participant engaged in two 30-minute RoboBlockly Studio sessions, corresponding to the \textit{AI--Only} and \textit{AI+Robot} conditions. The order of conditions was counterbalanced across AI--Only group and AI+Robot group. Within each session, students progressed through the three types of tasks while researchers conducted observations and video-recorded all interactions.

Following the task sessions, participants completed a 15-minute Bebras post-test, with four new problems matched in difficulty to the pre-test. After each condition, participants completed the NASA Task Load Index (NASA-TLX) to report perceived workload for that session. After completing both conditions, they filled out the System Usability Scale (SUS) and the User Experience Questionnaire (UEQ) to assess overall usability and experience with RoboBlockly Studio.

Finally, short semi-structured interviews were conducted to collect participants’ reflections on the tasks, challenges encountered, strategies used, and preferences regarding the system. Scratch paper and notes taken during tasks were collected to ensure completeness of study records.

\subsection{Analysis}
We conducted both quantitative and qualitative analyses to examine the impact of RoboBlockly Studio. 

\textbf{Quantitative.} \rebdel{Paired-sample \textit{t}-tests were used to compare pre- and post-test scores, with Wilcoxon signed-rank tests for non-parametric validation \cite{woolson2007wilcoxon,rosner2006wilcoxon}. Effect sizes (Cohen's $d_z$ for paired comparisons) were calculated. NASA-TLX results were summarized as means per dimension. SUS and UEQ were scored following their established protocols, with benchmark comparisons where appropriate.}\rebadd{Paired-sample \textit{t}-tests were used to compare pre- and post-test scores, with Wilcoxon signed-rank tests for non-parametric validation \cite{woolson2007wilcoxon,rosner2006wilcoxon}. Effect sizes (Cohen's $d_z$ for paired comparisons) were calculated. For NASA--TLX, we first assessed the normality of paired condition differences for each subscale and the Raw TLX score using Shapiro--Wilk tests. For variables with approximately normal paired differences, we report means and standard deviations and used paired-samples \textit{t}-tests. For variables that violated normality, we report medians and interquartile ranges and used Wilcoxon signed-rank tests. We report effect sizes as Cohen's $d_z$ for paired \textit{t}-tests and $r = Z/\sqrt{N}$ for Wilcoxon tests. SUS and UEQ were scored following their established protocols, with benchmark comparisons where appropriate.}

\textbf{Qualitative.} Semi-structured interview transcripts and video excerpts were analyzed using Reflexive Thematic Analysis (RTA)~\cite{braun2019reflecting}. Three coders independently identified themes, followed by iterative discussions to refine interpretations. We avoided deficit framing of learners and instead emphasized their agency, reflective strategies, and multimodal engagement.

\textbf{Triangulation.} Findings from tests, questionnaires, and qualitative data were cross-validated to build a coherent account of RoboBlockly Studio’s feasibility and impact \cite{denzin2007triangulation, heale2013understanding}.
Together, these procedures establish a robust user study design, ensuring both methodological rigor and alignment with HCI research practices.

\section{FINDINGS}\label{sec:findings}

Overall, the findings reveal that RoboBlockly Studio significantly improved students’ CT skills, provided engaging and scaffolded task experiences, and offered strong usability and user experience. Both quantitative measures (Bebras tests, NASA-TLX, SUS, UEQ) and qualitative insights (interviews, recordings) consistently support the system’s effectiveness. In line with prior HRI work such as PODER and Doodlebot \cite{cruz2025poder,williams2024doodlebot}, we also found that physical embodiment and multimodal interaction enhanced student engagement, while our results further \rebdel{demonstrate that the AI agent reduced cognitive workload and facilitated reflective debugging strategies.}\rebadd{suggest that the AI+Robot condition was associated with lower perceived workload and more reflective debugging strategies.}

\subsection{Performance in Bebras Pre- and Post-Tests}

\begin{figure}[t]
  \centering
  \includegraphics[width=0.6\linewidth]{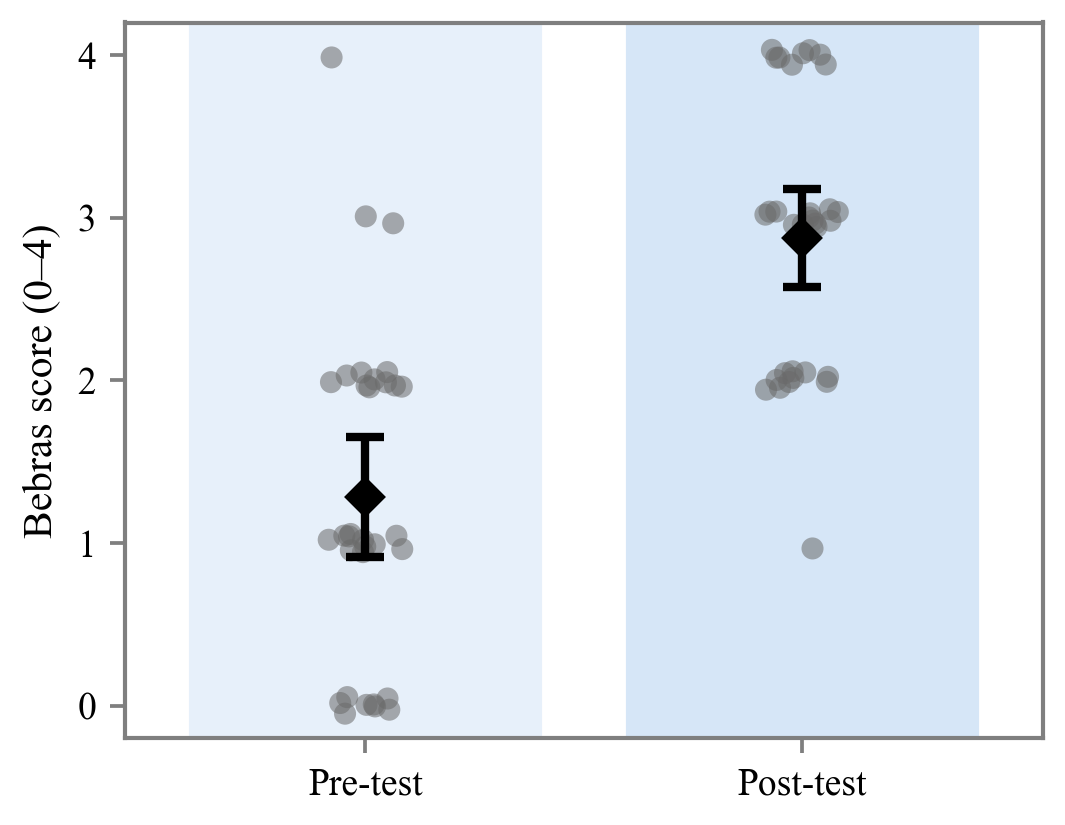}
  \caption{Bebras pre- and post-test scores (0--4). Diamonds indicate means with 95\% confidence intervals.}
  \Description{Dot-and-summary plot comparing Bebras scores (0–4) for Pre-test vs Post-test. Each condition shows multiple gray dots (individual students’ scores) and a black diamond marking the mean with vertical error bars for the 95\% confidence interval; the post-test distribution and mean are higher than the pre-test.}
  \label{fig:bebras_prepost}
\end{figure}

\textbf{Improvement trend.} Students’ computational thinking skills show\-ed a significant improvement from pre-test to post-test, and their problem-solving speed also increased. On average, participants solved 1.28 out of 4 tasks correctly in the pre-test (SD = 1.02), compared to 2.88 in the post-test (SD = 0.83). A paired-sample \textit{t}-test confirmed that this difference was statistically significant, $t(31)=9.53,\, p < .001$, with a large within-subject effect size ($d_z=1.69$). A Wilcoxon signed-rank test yielded consistent results ($W=430,\, p < .001$). These findings indicate that interacting with RoboBlockly Studio led to substantial gains in CT problem-solving performance (Fig.~\ref{fig:bebras_prepost}).

\textbf{Task difficulty effects.} Participants performed best on tasks of A and B difficulty levels, while accuracy was lower for C-level tasks. However, several students reported that after engaging with RoboBlockly Studio, they approached C-level problems in a more strategic way, demonstrating stronger decomposition, abstraction, and pattern recognition skills. For instance, P4 remarked: \textit{``In the pre-test I felt more pressed for time rather than finding the problems too difficult.''} P3 even requested to retry the pre-test item they had failed, saying: \textit{``I think I can solve it now.''}
These results address \rebdel{RQ1}\rebadd{RQ3}, showing measurable learning gains in computational thinking performance.

\subsection{Task Experience in RoboBlockly Studio}

Students completed three types of tasks in RoboBlockly Studio: completing missing blocks (introductory), fixing incorrect blocks (intermediate), and solving contextual problem scenarios (advanced). Across both conditions, students experienced the task progression as a structured learning trajectory that supported warm-up, sustained engagement, and increasingly reflective problem solving. Rather than treating tasks as isolated puzzles, many participants navigated them as iterative cycles of trying, checking, and revising, with the scaffolded progression shaping how they allocated attention and formed strategies.

\textbf{Progressive engagement.}
The introductory tasks typically served as an entry point where students quickly learned the interaction vocabulary of blocks (e.g., sequencing, loops, and function calls) and gained early confidence from visible progress. Several students described the progression from introductory to advanced tasks as motivating in itself: after finishing a missing-blocks task, they often immediately selected the next level “to see what’s harder.” For instance, P4 framed the experience as a staged challenge rather than a test: \textit{``It felt like a game---you clear the easy level first, then you level up to fight the boss.''} This perception of progression encouraged students to persist even when later tasks became demanding, because difficulty was interpreted as a natural next step rather than failure.

\textbf{From completion to diagnosis.}
We observed a consistent shift in how students reasoned across task types. In completing missing blocks (introductory), students primarily focused on local correctness---filling gaps and ensuring the program could run. Many students used trial-and-check behaviors (e.g., scanning block order, adjusting loop counts) to align the program with the expected outcome. As students moved to fixing incorrect blocks (intermediate), their behavior changed from “completing” to “diagnosing.” Students began to inspect programs as artifacts that could contain specific faults and to articulate what might be wrong before editing. For example, when encountering a suboptimal or flawed path, students often first pointed to the suspicious segment (e.g., an unnecessary detour or a misaligned turn) and then attempted small edits that could be quickly verified.

This diagnostic stance became most apparent in solving contextual problem scenarios (advanced), where tasks required planning over longer horizons and reasoning about constraints. Students frequently externalized their thinking by verbally rehearsing rules, breaking the scenario into subgoals, and proposing candidate strategies before committing to a complete solution. Several students described this shift as having to “think like the system” rather than simply placing blocks. As P7 explained after an advanced task: \textit{``I realized I had to think like a machine---step by step---otherwise I would miss a rule.''} Importantly, students did not reach this stance immediately; it emerged after they had experienced earlier tasks that established basic block literacy and expectations for iteration.

\textbf{Debugging as productive friction.}
The intermediate and advanced tasks elicited the richest moments of reflection, because they introduced purposeful errors or hidden constraints that were difficult to simulate mentally. When students’ initial solutions failed, many alternated between two concrete strategies: (1) revisiting the program structure (e.g., checking where a loop begins/ends, or whether a function encapsulates the intended repeated behavior), and (2) tracing the intended action sequence against the task constraints. In one common pattern, students first attempted broad edits (e.g., changing multiple blocks at once), then gradually narrowed down to single-step changes once they recognized that small modifications could isolate the fault. After receiving guidance that helped locate a specific mistake, students often expressed immediate relief and were able to explain what changed. For example, after the AI highlighted a faulty step in an intermediate task, P16 responded: \textit{``Oh---that’s exactly where it went wrong. I kept checking the end, but the mistake was earlier.''}

Overall, these observations indicate that the scaffolded task progression supported both engagement and a shift toward more systematic, reflective problem solving as task demands increased. These findings further support \rebdel{RQ1}\rebadd{RQ2} by showing that scaffolded task design promoted engagement and learning progression.

\subsection{Robot Embodiment as Feedback}
The counterbalanced within-subject design ensured that all participants experienced both versions of RoboBlockly Studio. Across conditions, we found that robot embodiment changed how students \emph{verified} and \emph{debugged} their programs: with a robot, learners treated execution as perceptual evidence to guide hypothesis-driven revisions; without a robot, learners relied more heavily on AI confirmation and on-screen inspection, which sometimes led to generic help-seeking and uncorrected logic errors.

\textbf{Embodiment as perceptual evidence.}
In the AI+Robot condition (RoboBlockly Studio), participants frequently initiated robot execution early, using the robot’s motion and interactions on the scenario mat as a concrete trace of program behavior. Rather than only asking whether a solution was ``correct,'' students often watched the run to localize where a failure emerged (e.g., an incorrect turn, a collision with an obstacle, or a missed target region), and then revised the corresponding block segment. This perceptual grounding also shaped how learners framed questions to the AI: after observing a discrepancy, students tended to ask about a specific step or block choice. For example, P1 initially struggled to distinguish turning from shifting and to encapsulate repeated actions into a function, but after consulting the AI with reference to the observed behavior, they reported rapid clarification: \textit{``I didn’t understand the difference between turning left and shifting left, or how to wrap actions into a function. But after asking the AI, I got it immediately.''} In our observations, such moments were often followed by learners verbalizing their reasoning (e.g., describing what the robot \emph{should} do next and which block change would produce that behavior), suggesting that embodiment supported reflective articulation rather than passive correction.

\textbf{Verification and debugging without embodiment.}
In the AI--Only condition, the conversational AI agent remained available to provide language-based hints, explanations, and guidance upon students’ request. However, without robot execution, learners verified program behavior primarily through on-screen inspection and AI dialogue. Some participants appreciated the reduced complexity of focusing narrowly on block logic. Yet others found the experience less interactive and more difficult to debug, particularly for longer, constraint-heavy problems where mental simulation was challenging. In these cases, help-seeking sometimes became generic and repetitive, with students requesting broad hints rather than identifying a concrete failure point. For instance, while working on the ``Wolf, Goat, and Cabbage'' task, P10 expressed frustration and a desire for more actionable guidance: \textit{``I really wish the AI could give me a more specific hint. I just can’t solve it.''} Consistent with this, we observed that several students in the AI--Only condition progressed further with partially incorrect reasoning, because there was no immediate, perceptually salient signal (e.g., a visible misstep or failed interaction) to prompt targeted revision.

\textbf{Learner preferences and perceived complementarity.}
\rebrep{After experiencing both conditions, most students preferred AI+Robot, describing it as more engaging and easier to reason about because the robot made success and failure visible in real time.}{After experiencing both conditions, \textbf{30 of 32} participants reported preferring AI+Robot, describing it as more engaging and easier to reason about because the robot made success and failure visible during execution.} Students often interpreted the robot run as a faster way to check assumptions, which reduced time spent guessing and increased confidence in revisions. At the same time, a minority of participants noted that the AI--Only version could be useful when they wanted to concentrate on block construction without attending to additional modalities, suggesting that the two modes can be complementary for different phases of learning (e.g., drafting logic versus validating behavior).

\rebrep{Overall, this comparison addresses RQ2 by showing that the AI agent and robot embodiment jointly enhanced engagement, debugging efficiency, and reflective reasoning. Similar to PODER, embodiment increased engagement \cite{cruz2025poder}, while our findings further suggest that embodied execution helps learners formulate more specific, block-grounded questions to the AI, supporting lower perceived workload during iterative debugging.}{Overall, these cross-condition observations primarily address RQ1 by suggesting that, within our study setting, robot execution supported more concrete verification and more targeted debugging questions. Similar to PODER, embodiment appeared to increase engagement \cite{cruz2025poder}, while our findings further suggest that embodied execution helped learners formulate more specific, block-grounded questions to the AI. Given the absence of richer screen-based baselines and direct task-level behavioral comparisons, we interpret this as suggestive rather than definitive evidence for the added value of embodiment.}

\subsection{Usability and Workload Assessments}
Participants completed the NASA-TLX after each condition; after completing both conditions, they completed the SUS and UEQ. Together, these measures provide a quantitative account of students’ perceived workload, usability, and overall experience, complementing the qualitative observations of how learners interacted with RoboBlockly Studio during iterative programming and debugging.

\begin{table*}[t]
  \centering
  \caption{\rebrep{NASA--TLX subscale scores (means and standard deviations) across conditions. Lower scores indicate a more favorable experience. We reverse-coded the TLX Performance subscale to maintain consistency with the other dimensions (i.e., lower Performance scores indicate higher perceived performance).}{NASA--TLX results across conditions. For subscales whose paired condition differences were approximately normal, descriptive statistics are reported as means and standard deviations and paired-samples \textit{t}-tests were used. For Frustration, which violated normality, descriptive statistics are reported as median and interquartile range and a Wilcoxon signed-rank test was used. Lower scores indicate a more favorable experience. For the standard NASA--TLX Performance subscale, lower scores indicate better self-rated task performance.}}
  \Description{Table reporting NASA--TLX results for AI--Only and AI+Robot. For Mental Demand, Physical Demand, Temporal Demand, Performance, Effort, and Raw TLX, descriptive statistics are shown as means and standard deviations. For Frustration, descriptive statistics are shown as median and interquartile range. The table also reports Shapiro--Wilk normality results on paired condition differences, the inferential test used, test statistics, p values, and effect sizes. Lower scores indicate a more favorable experience; for the standard NASA--TLX Performance subscale, lower scores indicate better self-rated task performance.}
  \label{tab:nasatlx}
  \scriptsize
  \setlength{\tabcolsep}{3pt}
  \begin{tabular}{@{}p{0.14\textwidth}ccccp{0.13\textwidth}p{0.11\textwidth}p{0.12\textwidth}cp{0.10\textwidth}@{}}
    \toprule
    & \multicolumn{2}{c}{\textbf{AI--Only}} & \multicolumn{2}{c}{\textbf{AI+Robot}} & \textbf{Shapiro--Wilk} & \textbf{Test} & \textbf{Statistic} & \textbf{$p$} & \textbf{Effect Size} \\
    \cmidrule(lr){2-3}\cmidrule(lr){4-5}
    \textbf{Subscale} & \textbf{$M$/Mdn} & \textbf{$SD$/IQR} & \textbf{$M$/Mdn} & \textbf{$SD$/IQR} &  &  &  &  &  \\
    \midrule
    Mental Demand
      & 12.83 & 4.02
      & 12.33 & 2.77
      & $W = .97,\ p = .45$
      & Paired \textit{t}-test
      & $t(31) = 0.73$
      & $.47$
      & $d_z = 0.13$ \\
    Physical Demand
      & 8.50 & 1.87
      & 5.17 & 3.51
      & $W = .95,\ p = .15$
      & Paired \textit{t}-test
      & $t(31) = 5.19$
      & $<.001^{***}$
      & $d_z = 0.92$ \\
    Temporal Demand
      & 13.67 & 2.73
      & 10.00 & 3.64
      & $W = .96,\ p = .35$
      & Paired \textit{t}-test
      & $t(31) = 5.41$
      & $<.001^{***}$
      & $d_z = 0.96$ \\
    Performance
      & 6.83 & 3.82
      & 6.08 & 4.29
      & $W = .97,\ p = .50$
      & Paired \textit{t}-test
      & $t(31) = 0.99$
      & $.33$
      & $d_z = 0.18$ \\
    Effort
      & 7.50 & 2.17
      & 4.92 & 3.42
      & $W = .95,\ p = .10$
      & Paired \textit{t}-test
      & $t(31) = 4.10$
      & $<.001^{***}$
      & $d_z = 0.72$ \\
    Frustration
      & 12 & 10--14
      & 2 & 1--3
      & $W = .85,\ p < .01$
      & Wilcoxon
      & $V = 528,\ Z = 4.94$
      & $<.001^{***}$
      & $r = 0.87$ \\
    \addlinespace[2pt]
    Raw TLX (unweighted mean)
      & 10.19 & 1.32
      & 6.77 & 1.79
      & $W = .96,\ p = .25$
      & Paired \textit{t}-test
      & $t(31) = 9.88$
      & $<.001^{***}$
      & $d_z = 1.75$ \\
    \bottomrule
  \end{tabular}
\end{table*}

\begin{figure}[t]
  \centering
  \includegraphics[width=\linewidth]{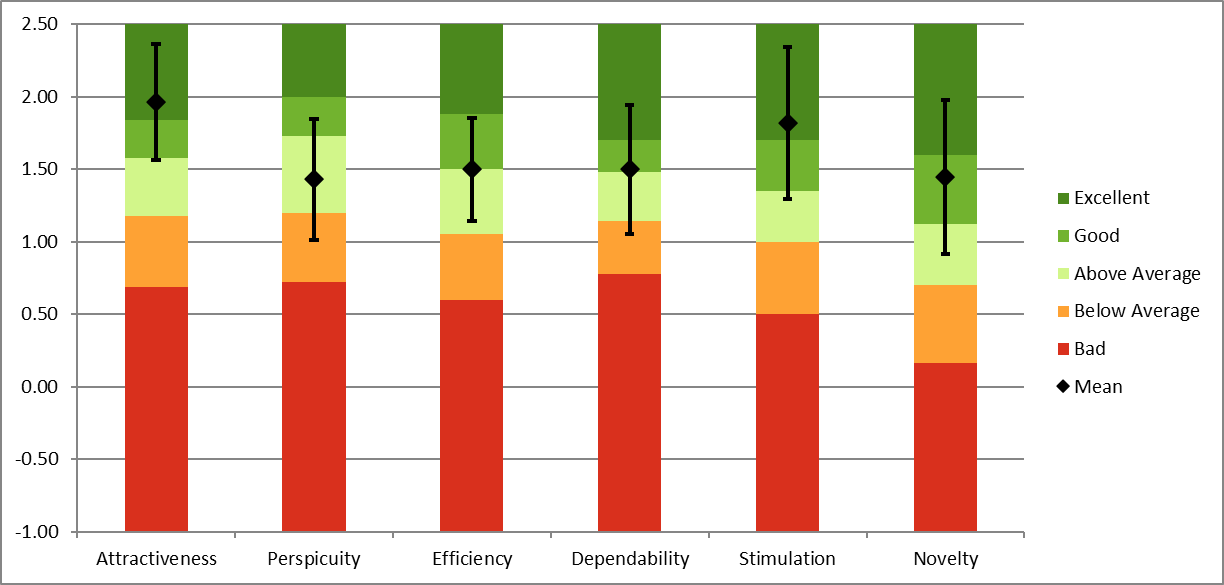}
  \caption{User Experience Questionnaire results for RoboBlockly Studio: mean and confidence intervals for the six UEQ dimensions with benchmark comparison.}
  \Description{Stacked bar chart of User Experience Questionnaire (UEQ) results for RoboBlockly Studio across six dimensions: Attractiveness, Perspicuity, Efficiency, Dependability, Stimulation, and Novelty, with colored bands indicating benchmark ranges from Bad to Excellent. Each dimension includes a black diamond for the mean and vertical error bars for the 95\% confidence interval; all six means are positive, with Attractiveness and Stimulation reaching the Excellent benchmark range and the remaining dimensions in Good or Above Average. }
  \label{fig:ueq_benchmark}
\end{figure}

\textbf{NASA--TLX.}
The NASA Task Load Index (NASA-TLX)~\cite{hart1988nasa} showed that participants experienced a moderate overall workload. Mental Demand remained relatively high in both conditions (AI--Only: $M=12.83$, AI+Robot: $M=12.33$), reflecting the inherent cognitive demands of reasoning about block-based programs, such as sequencing, loops, and constraints. Temporal Demand was also considerable, particularly in the AI--Only condition ($M=13.67$ vs.\ $M=10.00$ in AI+Robot), suggesting that the AI-augmented environment helped reduce time pressure during task completion. Physical Demand remained low across both conditions ($M=8.50$ vs.\ $M=5.17$), consistent with the classroom-based nature of the activity. \rebdel{After reverse-coding the Performance subscale so that higher values indicate higher workload,}\rebadd{For the standard NASA--TLX Performance subscale, lower scores indicate better self-rated task performance. In our data,} perceived performance-related workload was slightly lower in the AI+Robot condition ($M=6.08$) than in the AI--Only condition ($M=6.83$), suggesting that students felt more successful and less strained in accomplishing the tasks with AI+Robot support.

Notably, Effort and Frustration showed pronounced differences between conditions. Effort scores were lower in AI+Robot ($M=4.92$) compared to the AI--Only ($M=7.50$), indicating that AI support reduced the exertion required to make progress on the tasks. \rebdel{Frustration was markedly lower in AI+Robot ($M=2.17$) than in the AI--Only ($M=11.83$)}\rebadd{Frustration was markedly lower in the AI+Robot condition ($Mdn=2$, $IQR=1$--$3$) than in the AI--Only condition ($Mdn=12$, $IQR=10$--$14$). A Wilcoxon signed-rank test confirmed that this difference was significant ($V\approx 528$, $Z\approx 4.94$, $p<.001$, $r\approx .87$)}, aligning with interview comments that AI hints alleviated debugging difficulties. Interview data help contextualize these differences. For example, P6 explained that when encountering an error, \textit{``there was always something I could try next, either asking the AI or running the robot,''} which reduced the feeling of being stuck. Similarly, P14 described debugging with RoboBlockly as \textit{``less stressful, because I didn’t keep guessing blindly.''} These accounts suggest that conversational guidance and rapid verification supported sustained progress, mitigating the accumulation of time pressure and frustration observed in the AI--Only condition.

\textbf{SUS.}
The System Usability Scale (SUS) yielded an average score of $M = 72.5$, $SD = 8.3$, which falls within the \textit{Good Usability} range and exceeds the industry benchmark of 68 \cite{bangor2008empirical}. This indicates that RoboBlockly Studio was perceived as accessible and easy to use by novice learners. Participants frequently described the core workflow—selecting a task, assembling blocks, submitting code, and testing behavior—as intuitive and quick to grasp. For instance, P3 noted that \textit{``after a short time, I knew where to put blocks and how to test them.''} At the same time, a small number of students reported occasional difficulties in locating less frequently used features, such as specific blocks or interface controls. As P11 reflected, \textit{``most things were easy, but sometimes I had to search a bit for the right option.''} These comments are consistent with SUS results indicating generally good usability with a modest learning curve for advanced interactions.

\textbf{UEQ.}
The User Experience Questionnaire (UEQ) further captured students’ subjective experience with RoboBlockly Studio. As shown in Figure~\ref{fig:ueq_benchmark}, all six dimensions received positive evaluations (all $M > 1.3$, with 95\% CIs above the neutral threshold of $+0.8$). Relative to the standardized UEQ benchmark dataset, Attractiveness ($M=1.96$, 95\% CI [1.56, 2.36]) and Stimulation ($M=1.82$, 95\% CI [1.30, 2.34]) reached the \textit{Excellent} range, while Perspicuity ($M=1.43$) and Efficiency ($M=1.50$) were classified as \textit{Above Average}. Dependability ($M=1.50$) and Novelty ($M=1.44$) fell into the \textit{Good} category~\cite{hinderks2017design,laugwitz2008ueq}. These results suggest that students experienced the system as easy to learn, reliable, and motivating, with strong hedonic and pragmatic qualities. Participants’ comments echoed this pattern: P9 described the experience as \textit{``fun and clear at the same time,''} while P18 emphasized that \textit{``seeing things work made me want to keep trying.''}

Together, these questionnaire results provide evidence for RQ3, demonstrating that RoboBlockly Studio was perceived as usable, reliable, and positively experienced. Moreover, the inclusion of AI support contributed to reduced cognitive demand, highlighting its value in facilitating learning.

\section{DISCUSSION}

\subsection{Perceptual Feedback via Robot Embodiment}
In our design, we define embodied AI in practical terms as the integration of conversational guidance with robot embodiment, where the robot’s physical actions function as perceptual feedback for learners. Unlike purely virtual simulations, the robot’s embodied execution projects program logic into the physical world, making abstract constructs directly observable through spatial movement, object manipulation, and interaction with tangible constraints. This embodiment allows students to perceive not only whether a program runs but also how it unfolds in real space, thereby revealing both successes and errors in ways that are more immediate and perceptually salient. For example, when a student programs the robot to clean a tiled grid, the extent of physical coverage concretely demonstrates the adequacy of loop structures. Similarly, in the ancient ruin task where the robot must sequentially activate mechanisms, an erroneous path that crosses a collapsed area is manifested in the robot’s failed navigation. In both cases, the embodied performance delivers perceptual cues that anchor reasoning through tangible, real-world action.

\paragraph{Debugging with and without Embodiment.}
Learners in both conditions engaged in debugging, but their strategies differed. 
In the AI--Only condition, students often relied on explicit AI feedback, receiving direct indications of errors. 
In contrast, in the AI+Robot condition, learners primarily observed robot execution and used perceptual cues to identify errors before consulting the AI. 
Embodiment thus shifted debugging from receiving explicit messages to active perceptual monitoring and reasoning.

\paragraph{Verification and Error Correction.}
Verification of task completion also diverged between the two conditions.
In the AI--Only condition, learners typically relied on the AI’s evaluative feedback,
submitting their programs for judgment and receiving explicit confirmation or correction.
This approach provided a clear sense of accomplishment but positioned verification
primarily as an external process mediated by the AI.
In contrast, in the AI+Robot condition, learners often preferred to verify their solutions
by directly executing the robot and observing whether the physical task was completed as intended.
While the AI still offered supportive feedback, accomplishment in this condition was grounded
in embodied observation.

\paragraph{Challenging debugging scenarios.}
Certain subtle errors highlight differences in verification strategies between conditions.
For instance, in the wolf-goat-cabbage problem, learners must construct sequences of repeated
crossing and object manipulation actions, resulting in relatively long programs that are
difficult to mentally simulate in their entirety.
In the AI--Only condition, learners predominantly relied on direct AI queries, repeatedly
issuing simple prompts such as ``give a hint'' and updating their programs according to the
AI’s explicit guidance.
By contrast, in the AI+Robot condition, learners tended to first observe the robot’s execution,
reason about discrepancies, and independently identify potential errors before formulating
targeted queries to the AI.
This embodied feedback promoted active problem-solving and reflection, whereas the AI--Only
condition led to more reactive, prompt-driven debugging.
These differences underscore how embodiment influences learners’ engagement with verification
and error correction in complex, multi-step tasks.

\paragraph{Refined AI Queries after Observation.}
Following initial observation of the robot’s execution, learners in the AI+Robot condition
began formulating more specific and targeted queries to the AI.
Instead of relying on generic prompts, learners described precise discrepancies they had
noticed, asking for clarification or guidance on particular steps of their program.
In contrast, interactions in the AI--Only condition remained relatively generic and
hint-driven.
The combination of embodied observation and subsequent focused AI engagement allowed learners
to iteratively refine their programs with higher accuracy, demonstrating how embodiment
scaffolds more strategic and informed use of AI guidance in the debugging process.

\subsection{Visual Programming in Conversational Learning}
All participants engaged with RoboBlockly Studio’s hybrid environment, which combined block-based visual programming with conversational AI support. In this design, learners were not forced to alternate between separate systems for coding and asking questions. Instead, the workspace provided two tightly integrated modalities: (i) a visual canvas for constructing programs through drag-and-drop blocks, and (ii) a conversational panel for natural language interaction with the AI. Learners could submit partial or complete block sequences directly into the chat, preview execution, and immediately obtain AI guidance without leaving the programming context. This integration enabled learners to move fluidly between constructing logic visually and reasoning about it linguistically, thereby reducing cognitive fragmentation and fostering iterative refinement.

\paragraph{Block-AI Queries and Iterative Development.}
Beyond simply alternating between the two modalities, participants learned to strategically combine them. For example, participant P17 used the hybrid interface to send a partially constructed block sequence to the AI, accompanied by a natural language query: \textit{``I want to know if this is the right way to start the loop.''} The AI responded with targeted guidance, referencing the exact block configuration she had shared. This workflow demonstrated how learners externalized program structure visually while simultaneously verbalizing uncertainties through language. The ability to ``ask with blocks'' allowed participants to anchor their questions in concrete representations rather than abstract descriptions, which reduced ambiguity and improved the precision of AI responses. Such exchanges highlight how conversational interaction, when grounded in shared visual context, scaffolds iterative problem-solving and enables learners to progressively refine their solutions.

\paragraph{Accessible and Comprehensible AI Feedback.}
Participants also emphasized the clarity and accessibility of the AI’s feedback when integrated into the visual programming environment. For instance, P12 reported that the AI’s guidance was \textit{``simple and easy to understand,''} directly pointing out which portion of her block sequence required adjustment. This clarity was reinforced by the visual context: learners could immediately map the AI’s textual advice onto specific blocks in their workspace. Compared to traditional textual feedback alone, this coupling reduced confusion and minimized the cognitive effort needed to interpret suggestions. Quantitative results further supported these observations. NASA-TLX ratings indicated that learners in the AI+Robot condition experienced reduced temporal demand, lower effort, and much less frustration, consistent with reports that AI hints alleviated debugging difficulties. SUS scores placed the system above the established usability benchmark, confirming that participants found the hybrid environment accessible and easy to use. Moreover, positive UEQ ratings on perspicuity and efficiency aligned with participants’ reflections that AI feedback was clear and actionable. Together, these findings suggest that coupling conversational AI with visual program representation not only improved usability but also made learning interactions more comprehensible and cognitively manageable.

\subsection{Fostering Computational Thinking Skills}
\begin{figure*}[t]
  \centering
  \includegraphics[width=\linewidth]{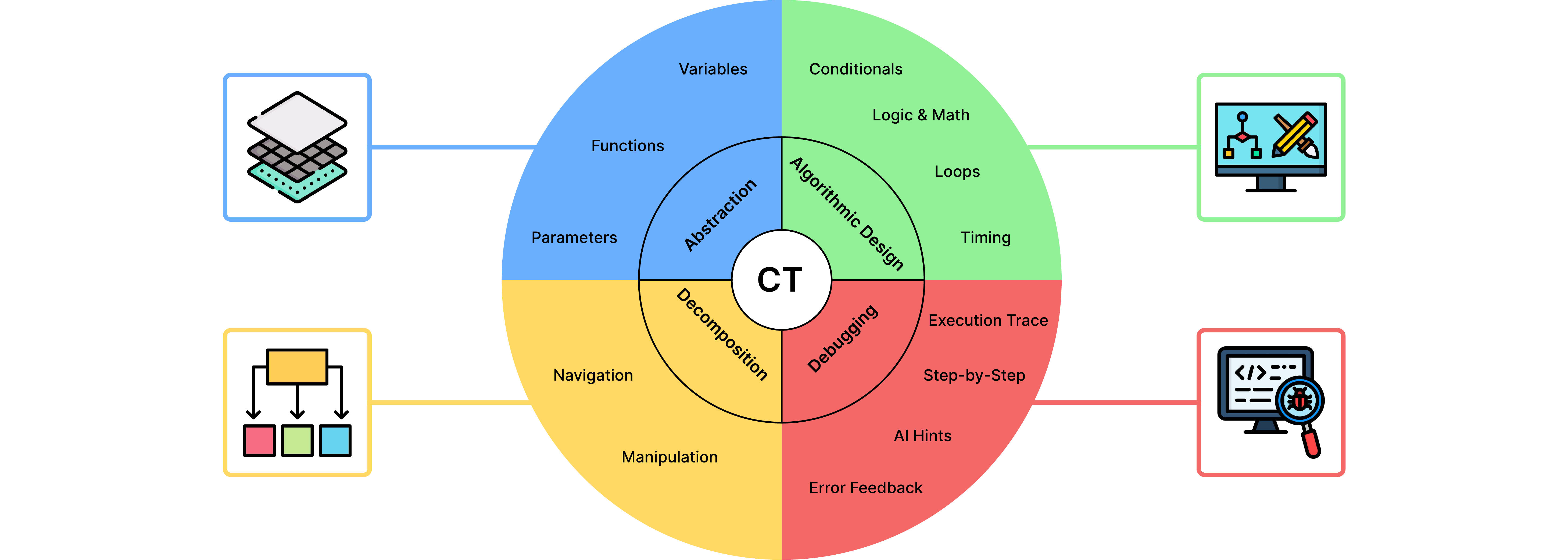}
  \caption{Mapping between RoboBlockly’s block categories, interaction supports, and core dimensions of computational thinking.}
  \Description{Concept map linking RoboBlockly features to computational thinking (CT). A central circle labeled CT is divided into four core pillars: Abstraction, Decomposition, Algorithmic Design, and Debugging, surrounded by example RoboBlockly elements: block categories that represent program structures (e.g., Variables, Functions, Conditionals, Logic \& Math, Loops, Parameters) and domain actions (Navigation, Manipulation, Timing), plus interaction supports for understanding and improving execution (Execution Trace, Step-by-Step, AI Hints, Error Feedback)}
  \label{fig:ct-mapping}
\end{figure*}

RoboBlockly Studio was designed to support computational thinking through the combination of task exercises, visual-conversational programming, and robot embodiment. As summarized in Fig.~\ref{fig:ct-mapping}, learners’ CT practices were grounded in (i) block categories that externalize key program structures (e.g., Variables, Functions, Conditionals, Logic \& Math, Loops, Parameters) and domain actions (e.g., Navigation, Manipulation, Timing), and (ii) interaction supports that make execution legible (Execution Trace, Step-by-Step) and provide guidance during iteration (AI Hints, Error Feedback). Together, these elements enabled learners to engage with core CT pillars---Abstraction, Decomposition, Algorithmic Design, and Debugging---in an integrated and contextually grounded manner.

\paragraph{Abstraction.}
RoboBlockly Studio supported abstraction by helping learners translate contextual constraints into reusable program concepts and parameterized representations rather than memorizing step-by-step actions. In tasks like the \textit{wolf-goat-cabbage river crossing} and \textit{knight’s chessboard} puzzle, students needed to reason about relational rules and movement constraints. They often used higher-level constructs (e.g., Functions and Parameters) to express repeated decision patterns, while AI-mediated conversational queries helped them clarify rules and intent without leaving the programming context. For example, when working on the river-crossing task, P8 initially attempted to enumerate individual moves but later encapsulated repeated crossings into a function, noting that \textit{``it’s the same rule each time, just with different things.''} Embodied execution further provided perceptual feedback about whether constraints were satisfied: when the robot violated a rule, students such as P15 revised their abstractions after observing the incorrect interaction, rather than adjusting isolated steps.

\paragraph{Decomposition.}
Learners broke down complex goals into manageable subproblems using both visual blocks and embodied execution. In the \textit{tile-cleaning} task, students supplemented missing loops in the cleaning function and decomposed the overall goal (cover the grid) into repeated routines and smaller action sequences. The robot’s traversal made it immediately apparent whether a decomposition was complete (e.g., whether a routine covered all required tiles), reinforcing the link between abstract subgoals and tangible outcomes. In several cases, students initially implemented long linear sequences, then reorganized them into smaller parts after observing redundant behavior. For instance, P4 paused execution midway and remarked that \textit{``this part keeps happening again,''} before extracting it into a separate routine, indicating a shift from surface-level completion toward structured decomposition.

\paragraph{Algorithmic Design.}
Across tasks, students constructed stepwise plans that satisfied logical constraints and, when applicable, physical execution requirements. For example, the \textit{minefield} challenge demanded energy-aware planning and efficient routing: learners needed to choose an action sequence that balanced exploration with constraints such as limited energy and obstacles. Students frequently moved between planning and implementation by sketching a route, encoding it with Navigation/Timing/Logic \& Math blocks, and then refining the algorithm when execution revealed inefficiencies or constraint violations. P12, for instance, adjusted their route after noticing that an early detour caused energy depletion later in the run, explaining that \textit{``the order matters, not just the steps.''} In this way, algorithmic design was practiced as an iterative process of proposing, implementing, and optimizing strategies within the task’s rule structure.

\paragraph{Debugging.}
Debugging was foregrounded by tasks that contained deliberate flaws or subtle constraints, and was supported by system features that made program behavior inspectable and revisable. In the \textit{ancient ruin} and \textit{minefield} tasks, students often used perceptual cues from robot execution to localize failures (e.g., an incorrect turn, an obstacle collision, or an unintended detour), then adjusted the corresponding blocks. For example, after observing the robot collide with an obstacle at a specific corner, P10 traced the execution step-by-step and corrected a misplaced turn block, stating that \textit{``it’s wrong right here, not at the end.''} When embodiment was not available, learners relied more on stepwise inspection and AI feedback to identify where reasoning diverged from expected behavior. Interaction supports such as Execution Trace and Step-by-Step helped learners connect observed outcomes to specific program segments, while AI Hints and Error Feedback supported targeted revisions. Together, these supports encouraged learners to treat errors as actionable signals for refining both their code and their underlying strategy.

\subsection{Limitations and Future Work}
\rebdel{While our study highlights the potential of RoboBlockly Studio in supporting novice learners’ computational thinking, several limitations should be acknowledged. First, the user study involved a relatively small sample size which may limit the generalizability of our findings. Future studies with larger and more diverse populations are needed to validate the robustness of our results. Second, the short duration of the study primarily captured immediate usability and learning outcomes; longer-term deployments are required to examine how RoboBlockly Studio impacts sustained learning and retention of computational thinking concepts. Third, the study did not include a traditional control group, such as students solving CT tasks without AI support. Another limitation lies in the controlled study environment, which may not fully reflect the variability of classroom dynamics.}

\rebadd{While our study highlights the potential of RoboBlockly Studio in supporting novice learners’ computational thinking, several limitations should be acknowledged. First, the user study involved a relatively small sample, which limits the generalizability of our findings. Second, the short duration of the study primarily captured immediate usability and learning outcomes; longer-term deployments are needed to examine sustained learning and retention of CT concepts. Third, although the study compared AI-only and AI+Robot conditions, it did not include richer screen-based baselines such as animated execution traces or step-by-step simulation, nor did it clearly distinguish CT gains between the two conditions. Fourth, although the LLM-based checking was generally reliable for standard task solutions, it occasionally incorrectly flagged functionally correct but unconventional block sequences as incorrect. Finally, the study took place in a controlled classroom environment, which may not fully reflect the variability of everyday classroom dynamics.}

Future work will explore expanding the range of supported tasks, \rebadd{improving the robustness of LLM-based checking for diverse valid solution strategies,} integrating adaptive difficulty adjustment, and investigating collaborative modes where multiple learners interact with the AI tutor and robot together. \rebadd{Comparisons with richer screen-based baselines are needed to better contextualize the role of physical embodiment. Longer-term deployments with condition-level analyses are also needed to examine sustained CT development and to clarify differences in CT gains between the AI-only and AI+Robot conditions.} We also aim to incorporate richer analytics for teachers to monitor student progress and difficulties in real time. Finally, investigating the scalability of RoboBlockly Studio in formal classroom curricula, as well as its potential integration with other AI-driven learning platforms, represents an important direction for enhancing its educational impact.

\section{CONCLUSION}
RoboBlockly Studio demonstrates how conversational tutoring, block-based programming, and robot execution can be combined to help novices see, discuss, and revise computation as embodied action within a single interactive system. In a within-subject study, participants showed significant gains on CT tasks, reported lower workload—especially in temporal demand, effort, and frustration—and rated the system as both usable and engaging. \rebrep{Embodiment acted as perceptual feedback, making successes and errors visible in real time, which encouraged learners to verify solutions through observation and to ask the AI more targeted questions.}{Embodied execution functioned as a perceptual feedback channel that made successes and errors more inspectable during task execution, which in turn appeared to encourage learners to verify solutions through observation and to ask the AI more targeted questions.} In addition, the conversational agent reduced time pressure and frustration while preserving agency by allowing students to decide when to request hints or test their programs on the robot.

\begin{acks}
The authors would like to sincerely thank Gusto Wu and the teachers from the STEAM Center at Ulink High School of Suzhou Industrial Park, Yoga Lu from Suzhou Industrial Park Foreign Language School, and Zhang Chao from the Robotics Club Center of Jiangyin No. 1 High School, Jiangsu Province, for their valuable support in facilitating the implementation of this research. This study was supported by Xi'an Jiaotong-Liverpool University (Grant No. RDF-23-02-082 and RDF-22-01-062).
\end{acks}

\section*{Generative AI Usage Disclosure}

The system described in this paper uses gpt-4o as the core LLM for our conversational agent within the application. GPT 4.0 was used to assist with non-novel boilerplate code, as well as to support general application development. All core logic and system components were designed and implemented by the authors. All AI-assisted code \rebdel{were}\rebadd{was} reviewed, tested, and edited by the authors to ensure factual accuracy, correctness, and reproducibility, in accordance with ACM’s Generative AI usage policy.

\bibliographystyle{ACM-Reference-Format}
\bibliography{references}

@String{Computing = "Computing" }

@String{Computer = "{IEEE} Computer" }

@String{Springer = "Springer-Verlag" }

@BOOK{test,
   author = "Donald E. Knuth",
   title = "Seminumerical Algorithms",
   volume = 2,
   series = "The Art of Computer Programming",
   publisher = "Addison-Wesley",
   address = "Reading, MA",
   edition = "2nd",
   month = "10~" # jan,
   year = "1981",
}

@article{sung2017introducing,
  title={Introducing computational thinking to young learners: Practicing computational perspectives through embodiment in mathematics education},
  author={Sung, Woonhee and Ahn, Junghyun and Black, John B},
  journal={Technology, Knowledge and Learning},
  volume={22},
  number={3},
  pages={443--463},
  year={2017},
  publisher={Springer}
}

@article{resnick2009scratch,
  title={Scratch: programming for all},
  author={Resnick, Mitchel and Maloney, John and Monroy-Hern{\'a}ndez, Andr{\'e}s and Rusk, Natalie and Eastmond, Evelyn and Brennan, Karen and Millner, Amon and Rosenbaum, Eric and Silver, Jay and Silverman, Brian and others},
  journal={Communications of the ACM},
  volume={52},
  number={11},
  pages={60--67},
  year={2009},
  publisher={ACM New York, NY, USA}
}

@article{bangor2008empirical,
  title={An empirical evaluation of the system usability scale},
  author={Bangor, Aaron and Kortum, Philip T and Miller, James T},
  journal={Intl. Journal of Human--Computer Interaction},
  volume={24},
  number={6},
  pages={574--594},
  year={2008},
  publisher={Taylor \& Francis}
}

@article{cao2009nasa,
  title={NASA TLX: Software for assessing subjective mental workload},
  author={Cao, Alex and Chintamani, Keshav K and Pandya, Abhilash K and Ellis, R Darin},
  journal={Behavior research methods},
  volume={41},
  number={1},
  pages={113--117},
  year={2009},
  publisher={Springer}
}

@incollection{hart1988nasa,
  title        = {Development of NASA-TLX (Task Load Index): Results of Empirical and Theoretical Research},
  author       = {Hart, Sandra G. and Staveland, Lowell E.},
  booktitle    = {Advances in Psychology},
  volume       = {52},
  pages        = {139--183},
  year         = {1988},
  publisher    = {Elsevier},
  doi          = {10.1016/S0166-4115(08)62386-9}
}

@article{lewis2018system,
  title={The system usability scale: past, present, and future},
  author={Lewis, James R},
  journal={International Journal of Human--Computer Interaction},
  volume={34},
  number={7},
  pages={577--590},
  year={2018},
  publisher={Taylor \& Francis}
}

@article{hinderks2017design,
  title={Design and evaluation of a short version of the user experience questionnaire (UEQ-S)},
  author={Hinderks, Andreas},
  journal={International Journal of Interactive Multimedia and Artificial Intelligence},
  year={2017}
}

@article{adeoye2021research,
  title={Research and scholarly methods: Semi-structured interviews},
  author={Adeoye-Olatunde, Omolola A and Olenik, Nicole L},
  journal={Journal of the american college of clinical pharmacy},
  volume={4},
  number={10},
  pages={1358--1367},
  year={2021},
  publisher={Wiley Online Library}
}

@inproceedings{dagiene2008bebras,
  title={Bebras international contest on informatics and computer literacy: Criteria for good tasks},
  author={Dagien{\.e}, Valentina and Futschek, Gerald},
  booktitle={International conference on informatics in secondary schools-evolution and perspectives},
  pages={19--30},
  year={2008},
  organization={Springer}
}

@inproceedings{dagiene2016s,
  title={It’s computational thinking! Bebras tasks in the curriculum},
  author={Dagien{\.e}, Valentina and Sentance, Sue},
  booktitle={International conference on informatics in schools: Situation, evolution, and perspectives},
  pages={28--39},
  year={2016},
  organization={Springer}
}

@article{braun2019reflecting,
  title={Reflecting on reflexive thematic analysis},
  author={Braun, Virginia and Clarke, Victoria},
  journal={Qualitative research in sport, exercise and health},
  volume={11},
  number={4},
  pages={589--597},
  year={2019},
  publisher={Taylor \& Francis}
}

@article{heale2013understanding,
  title={Understanding triangulation in research},
  author={Heale, Roberta and Forbes, Dorothy},
  journal={Evidence-based nursing},
  volume={16},
  number={4},
  pages={98--98},
  year={2013},
  publisher={Royal College of Nursing}
}

@article{denzin2007triangulation,
  title={Triangulation},
  author={Denzin, Norman K},
  journal={The Blackwell encyclopedia of sociology},
  year={2007},
  publisher={Wiley Online Library}
}

@article{woolson2007wilcoxon,
  title={Wilcoxon signed-rank test},
  author={Woolson, Robert F},
  journal={Wiley encyclopedia of clinical trials},
  pages={1--3},
  year={2007},
  publisher={Wiley Online Library}
}

@article{rosner2006wilcoxon,
  title={The Wilcoxon signed rank test for paired comparisons of clustered data},
  author={Rosner, Bernard and Glynn, Robert J and Lee, Mei-Ling T},
  journal={Biometrics},
  volume={62},
  number={1},
  pages={185--192},
  year={2006},
  publisher={Oxford University Press}
}

@inproceedings{cruz2025poder,
  title={PODER: A Robot Programming Framework to Further Inclusion of People with Mild Cognitive Impairment in HRI Research},
  author={Cruz-Sandoval, Dagoberto and Murakami, Michele and Kubota, Alyssa and Riek, Laurel D},
  booktitle={2025 20th ACM/IEEE International Conference on Human-Robot Interaction (HRI)},
  pages={599--609},
  year={2025},
  organization={IEEE}
}

@inproceedings{williams2024doodlebot,
  title={Doodlebot: An educational robot for creativity and AI literacy},
  author={Williams, Randi and Ali, Safinah and Alcantara, Ra{\'u}l and Burghleh, Tasneem and Alghowinem, Sharifa and Breazeal, Cynthia},
  booktitle={Proceedings of the 2024 ACM/IEEE international conference on human-robot interaction},
  pages={772--780},
  year={2024}
}

@article{noordin2025computational,
  author    = {Noordin, Nurul Hazlina},
  title     = {Computational Thinking Through Scaffolded Game Development Activities: A Study with Graphical Programming},
  journal   = {European Journal of Educational Research},
  volume    = {14},
  number    = {4},
  pages     = {1137--1149},
  year      = {2025},
  publisher = {Eurasian Society of Educational Research},
  doi       = {10.12973/eu-jer.14.4.1137},
  url       = {https://doi.org/10.12973/eu-jer.14.4.1137}
}

@article{code2020agency,
  title={Agency for learning: Intention, motivation, self-efficacy and self-regulation},
  author={Code, Jillianne},
  journal={Frontiers in Education},
  volume={5},
  pages={19},
  year={2020},
  publisher={Frontiers Media SA}
}

@article{fennell2019computational,
  title={Computational apprenticeship: Cognitive apprenticeship for the digital era},
  author={Fennell, Hayden and Lyon, Joseph A and Madamanchi, Aasakiran and Magana, Alejandra J},
  journal={Journal of Engineering Education},
  volume={109},
  number={170.176},
  year={2019}
}

@inproceedings{lee2012ctarcade,
  title        = {CTArcade: Learning Computational Thinking While Training Virtual Characters Through Game Play},
  author       = {Lee, Tak Yeon and Mauriello, Matthew Louis and Ingraham, John and Sopan, Awalin and Ahn, June and Bederson, Benjamin B.},
  booktitle    = {CHI '12 Extended Abstracts on Human Factors in Computing Systems},
  pages        = {2309--2314},
  year         = {2012},
  publisher    = {ACM}
}

@inproceedings{rowe2018labeling,
  title={Labeling implicit computational thinking in pizza pass gameplay},
  author={Rowe, Elizabeth and Asbell-Clarke, Jodi and Baker, Ryan and Gasca, Santiago and Bardar, Erin and Scruggs, Richard},
  booktitle={Extended abstracts of the 2018 CHI conference on human factors in computing systems},
  pages={1--6},
  year={2018}
}

@inproceedings{zaman2023exploring,
  title={Exploring Computational Thinking Practices and Gestures in the Context of Matrix Math},
  author={Zaman, Ulia},
  booktitle={Extended Abstracts of the 2023 CHI Conference on Human Factors in Computing Systems},
  pages={1--6},
  year={2023}
}

@inproceedings{das2025cultivating,
  title={Cultivating Computational Thinking and Social Play among Neurodiverse Preschoolers in Inclusive Classrooms},
  author={Das, Maitraye and Tran, Megan and Ong, Amanda Chih-han and Kientz, Julie A and Feldner, Heather},
  booktitle={Proceedings of the 2025 CHI Conference on Human Factors in Computing Systems},
  pages={1--22},
  year={2025}
}

@inproceedings{troiano2025ct4all,
  title={CT4ALL: Towards Putting Teachers in the Loop to Advance Automated Computational Thinking Metric Assessments in Game-Based Learning},
  author={Troiano, Giovanni M and Cassidy, Michael and Morales, Daniel Escobar and Pons, Guillermo and Abdollahi, Amir and Robles, Gregorio and Puttick, Gillian and Harteveld, Casper},
  booktitle={Proceedings of the 2025 CHI Conference on Human Factors in Computing Systems},
  pages={1--23},
  year={2025}
}

@article{kafai2022revaluation,
  title={A revaluation of computational thinking in K--12 education: Moving toward computational literacies},
  author={Kafai, Yasmin B and Proctor, Chris},
  journal={Educational Researcher},
  volume={51},
  number={2},
  pages={146--151},
  year={2022},
  publisher={SAGE Publications Sage CA: Los Angeles, CA}
}

@inproceedings{dong2019prada,
  title={PRADA: A practical model for integrating computational thinking in K-12 education},
  author={Dong, Yihuan and Catete, Veronica and Jocius, Robin and Lytle, Nicholas and Barnes, Tiffany and Albert, Jennifer and Joshi, Deepti and Robinson, Richard and Andrews, Ashley},
  booktitle={Proceedings of the 50th ACM technical symposium on computer science education},
  pages={906--912},
  year={2019}
}

@inproceedings{laugwitz2008ueq,
  title        = {Construction and Evaluation of a User Experience Questionnaire},
  author       = {Laugwitz, Bettina and Held, Theo and Schrepp, Martin},
  booktitle    = {HCI and Usability for Education and Work (USAB 2008)},
  series       = {Lecture Notes in Computer Science},
  volume       = {5298},
  pages        = {63--76},
  year         = {2008},
  publisher    = {Springer}
}

@article{weintrop2016defining,
  title={Defining computational thinking for mathematics and science classrooms},
  author={Weintrop, David and Beheshti, Elham and Horn, Michael and Orton, Kai and Jona, Kemi and Trouille, Laura and Wilensky, Uri},
  journal={Journal of science education and technology},
  volume={25},
  number={1},
  pages={127--147},
  year={2016},
  publisher={Springer}
}

@article{lane2023teacher,
  title={Teacher re-novicing on the path to integrating computational thinking in high school physics instruction},
  author={Lane, W Brian and Galanti, Terrie M and Rozas, XL},
  journal={Journal for STEM Education Research},
  volume={6},
  number={2},
  pages={302--325},
  year={2023},
  publisher={Springer}
}

@article{tagare2024factors,
  title={Factors That Predict K-12 Teachers' Ability to Apply Computational Thinking Skills},
  author={Tagare, Deepti},
  journal={ACM Transactions on Computing Education},
  volume={24},
  number={1},
  pages={1--26},
  year={2024},
  publisher={ACM New York, NY}
}

@article{ouyang2024effects,
  title={The effects of educational robotics in STEM education: A multilevel meta-analysis},
  author={Ouyang, Fan and Xu, Weiqi},
  journal={International Journal of STEM education},
  volume={11},
  number={1},
  pages={7},
  year={2024},
  publisher={Springer}
}

@inproceedings{dietz2023visual,
  title={Visual storycoder: A multimodal programming environment for children’s creation of stories},
  author={Dietz, Griffin and Tamer, Nadin and Ly, Carina and Le, Jimmy K and Landay, James A},
  booktitle={Proceedings of the 2023 CHI Conference on Human Factors in Computing Systems},
  pages={1--16},
  year={2023}
}

@inproceedings{chen2024chatscratch,
  title={ChatScratch: An AI-augmented system toward autonomous visual programming learning for children aged 6-12},
  author={Chen, Liuqing and Xiao, Shuhong and Chen, Yunnong and Song, Yaxuan and Wu, Ruoyu and Sun, Lingyun},
  booktitle={Proceedings of the 2024 CHI Conference on Human Factors in Computing Systems},
  pages={1--19},
  year={2024}
}

@article{robe2022designing,
  title={Designing pairbuddy—a conversational agent for pair programming},
  author={Robe, Peter and Kuttal, Sandeep Kaur},
  journal={ACM Transactions on Computer-Human Interaction (TOCHI)},
  volume={29},
  number={4},
  pages={1--44},
  year={2022},
  publisher={ACM New York, NY}
}

@inproceedings{kazemitabaar2024codeaid,
  title={Codeaid: Evaluating a classroom deployment of an llm-based programming assistant that balances student and educator needs},
  author={Kazemitabaar, Majeed and Ye, Runlong and Wang, Xiaoning and Henley, Austin Zachary and Denny, Paul and Craig, Michelle and Grossman, Tovi},
  booktitle={Proceedings of the 2024 chi conference on human factors in computing systems},
  pages={1--20},
  year={2024}
}

@article{li2025visualcodemooc,
  title={VisualCodeMOOC: A course platform for algorithms and data structures integrating a conversational agent for enhanced learning through dynamic visualizations},
  author={Li, Mingyuan and Wang, Duan and Purwanto, Erick and Selig, Thomas and Zhang, Qing and Liang, Hai-Ning},
  journal={SoftwareX},
  volume={30},
  pages={102072},
  year={2025},
  publisher={Elsevier}
}

@inproceedings{pan2025tutorup,
  title={Tutorup: What if your students were simulated? training tutors to address engagement challenges in online learning},
  author={Pan, Sitong and Schmucker, Robin and Garcia Bulle Bueno, Bernardo and Llanes, Salome Aguilar and Albo Alarc{\'o}n, Fernanda and Zhu, Hangxiao and Teo, Adam and Xia, Meng},
  booktitle={Proceedings of the 2025 CHI Conference on Human Factors in Computing Systems},
  pages={1--18},
  year={2025}
}

@article{liu2024bringing,
  title     = {Bringing computational thinking into classrooms: a systematic review on supporting teachers in integrating computational thinking into K-12 classrooms},
  author    = {Liu, Zhichun and Gearty, Zarina and Richard, Eleanor and Orrill, Chandra Hawley and Kayumova, Shakhnoza and Balasubramanian, Ramprasad},
  journal   = {International Journal of STEM Education},
  volume    = {11},
  number    = {},
  pages     = {51},
  year      = {2024},
  publisher = {SpringerOpen}
}

@article{chen2024pbl,
  title     = {Put two and two together: A systematic review of combining computational thinking and project-based learning in STEM classrooms},
  author    = {Chen, Jiaxin and Hui, Jiaojiao},
  journal   = {STEM Education Review},
  volume    = {2},
  pages     = {1--18},
  year      = {2024},
  publisher = {Hong Kong Society of Medical Publishing}
}

@article{huang2021unplugged,
  title     = {A critical review of literature on ``unplugged'' pedagogies in K--12 computer science and computational thinking education},
  author    = {Huang, Wendy and Looi, Chee-Kit},
  journal   = {Computer Science Education},
  volume    = {31},
  number    = {1},
  pages     = {83--111},
  year      = {2021},
  publisher = {Taylor \& Francis}
}

@inproceedings{limke2023empowering,
  title={Empowering students as leaders of co-design for block-based programming},
  author={Limke, Ally and Lytle, Nicholas and Lin, Maggie and Mahmoud, Sana and Hill, Marnie and Catet{\'e}, Veronica and Barnes, Tiffany},
  booktitle={Extended Abstracts of the 2023 CHI Conference on Human Factors in Computing Systems},
  pages={1--7},
  year={2023}
}

@inproceedings{fronchetti2024block,
  title={Block-based programming for two-armed robots: a comparative study},
  author={Fronchetti, Felipe and Ritschel, Nico and Schorr, Logan and Barfield, Chandler and Chang, Gabriella and Spinola, Rodrigo and Holmes, Reid and Shepherd, David C},
  booktitle={Proceedings of the 46th IEEE/ACM International Conference on Software Engineering},
  pages={1--12},
  year={2024}
}

@inproceedings{rocha2025awareness,
  title={Awareness in Collaborative Mixed-Visual Ability Tangible Programming Activities},
  author={Rocha, Filipa and Sim{\~a}o, Hugo and Nogueira, Jo{\~a}o and Neto, Isabel and Guerreiro, Tiago and Nicolau, Hugo},
  booktitle={Proceedings of the 2025 CHI Conference on Human Factors in Computing Systems},
  pages={1--15},
  year={2025}
}

@inproceedings{zhou2025instructpipe,
  title={InstructPipe: Generating Visual Blocks Pipelines with Human Instructions and LLMs},
  author={Zhou, Zhongyi and Jin, Jing and Phadnis, Vrushank and Yuan, Xiuxiu and Jiang, Jun and Qian, Xun and Wright, Kristen and Sherwood, Mark and Mayes, Jason and Zhou, Jingtao and others},
  booktitle={Proceedings of the 2025 CHI Conference on Human Factors in Computing Systems},
  pages={1--22},
  year={2025}
}

@inproceedings{williams2019artificial,
  title={A is for artificial intelligence: the impact of artificial intelligence activities on young children's perceptions of robots},
  author={Williams, Randi and Park, Hae Won and Breazeal, Cynthia},
  booktitle={Proceedings of the 2019 CHI conference on human factors in computing systems},
  pages={1--11},
  year={2019}
}

@inproceedings{pedersen2021educational,
  title={Educational robotics and mediated transfer: transitioning from tangible tile-based programming, to visual block-based programming},
  author={Pedersen, Bjarke Kristian Maigaard Kj{\ae}r and Jacobsen, Didde Marie and Teichert, Lukas Juhl Lyk and Nielsen, Jacob},
  booktitle={Companion of the 2021 ACM/IEEE International Conference on Human-Robot Interaction},
  pages={402--406},
  year={2021}
}

@inproceedings{chi2024interactive,
  title={Interactive human-robot teaching recovers and builds trust, even with imperfect learners},
  author={Chi, Vivienne Bihe and Malle, Bertram F},
  booktitle={Proceedings of the 2024 ACM/IEEE International Conference on Human-Robot Interaction},
  pages={127--136},
  year={2024}
}

@inproceedings{booth2022revisiting,
  title={Revisiting human-robot teaching and learning through the lens of human concept learning},
  author={Booth, Serena and Sharma, Sanjana and Chung, Sarah and Shah, Julie and Glassman, Elena L},
  booktitle={2022 17th ACM/IEEE International Conference on Human-Robot Interaction (HRI)},
  pages={147--156},
  year={2022},
  organization={IEEE}
}

@inproceedings{lewis2025physiobots,
  author       = {Myles Lewis and Pranay Joshi and Wesley Cade Junkins and Vincent Ingram and Chris S. Crawford},
  title        = {PhysioBots: Engaging K-12 Students with Physiological Computing and Robotics},
  booktitle    = {Extended Abstracts of the 2025 CHI Conference on Human Factors in Computing Systems (CHI EA '25)},
  year         = {2025},
  publisher    = {Association for Computing Machinery},
  address      = {New York, NY, USA},
  pages        = {1--8},
  articleno    = {434},
  url          = {https://doi.org/10.1145/3706599.3720106}
}

@article{grover2013ct,
  author    = {Shuchi Grover and Roy Pea},
  title     = {Computational Thinking in K--12: A Review of the State of the Field},
  journal   = {Educational Researcher},
  volume    = {42},
  number    = {1},
  pages     = {38--43},
  year      = {2013},
  publisher = {SAGE Publications}
}

@techreport{wing2010ct,
  author       = {Wing, Jeannette M.},
  title        = {Computational Thinking: What and Why?},
  institution  = {Carnegie Mellon University, School of Computer Science},
  year         = {2010},
  month        = nov,
  note         = {17 November 2010},
  url          = {https://www.cs.cmu.edu/~CompThink/resources/TheLinkWing.pdf}
}

@article{SHUTE2017142,
title = {Demystifying computational thinking},
journal = {Educational Research Review},
volume = {22},
pages = {142-158},
year = {2017},
issn = {1747-938X},
doi = {https://doi.org/10.1016/j.edurev.2017.09.003},
url = {https://www.sciencedirect.com/science/article/pii/S1747938X17300350},
author = {Valerie J. Shute and Chen Sun and Jodi Asbell-Clarke}
}

@inproceedings{10.1145/3357236.3395497,
  author    = {Sellier, Nine and An, Pengcheng},
  title     = {How Peripheral Interactive Systems Can Support Teachers with Differentiated Instruction: Using FireFlies as a Probe},
  booktitle = {Proceedings of the 2020 ACM Designing Interactive Systems Conference},
  series    = {DIS '20},
  year      = {2020},
  isbn      = {9781450369749},
  location  = {Eindhoven, Netherlands},
  pages     = {1117--1129},
  numpages  = {13},
  url       = {https://doi.org/10.1145/3357236.3395497},
  doi       = {10.1145/3357236.3395497},
  acmid     = {3395497},
  publisher = {ACM},
  address   = {New York, NY, USA}
}

@inproceedings{10.1145/3322276.3322365,
  author    = {d'Anjou, Bernice and Bakker, Saskia and An, Pengcheng and Bekker, Tilde},
  title     = {How Peripheral Data Visualisation Systems Support Secondary School Teachers during VLE-Supported Lessons},
  booktitle = {Proceedings of the 2019 on Designing Interactive Systems Conference},
  series    = {DIS '19},
  year      = {2019},
  isbn      = {9781450358507},
  location  = {San Diego, CA, USA},
  pages     = {859--870},
  numpages  = {12},
  url       = {https://doi.org/10.1145/3322276.3322365},
  doi       = {10.1145/3322276.3322365},
  acmid     = {3322365},
  publisher = {ACM},
  address   = {New York, NY, USA}
}

@inproceedings{10.1145/3461778.3462084,
  author    = {Lu, Alex Jiahong and Marcu, Gabriela and Ackerman, Mark S. and Dillahunt, Tawanna R},
  title     = {Coding Bias in the Use of Behavior Management Technologies: Uncovering Socio-technical Consequences of Data-driven Surveillance in Classrooms},
  booktitle = {Proceedings of the 2021 ACM Designing Interactive Systems Conference},
  series    = {DIS '21},
  year      = {2021},
  isbn      = {9781450384766},
  location  = {Virtual Event, USA},
  pages     = {508--522},
  numpages  = {15},
  url       = {https://doi.org/10.1145/3461778.3462084},
  doi       = {10.1145/3461778.3462084},
  acmid     = {3462084},
  publisher = {ACM},
  address   = {New York, NY, USA}
}

@inproceedings{10.1145/3563657.3596079,
  author    = {Ngoon, Tricia J. and Kovalev, David and Patidar, Prasoon and Harrison, Chris and Agarwal, Yuvraj and Zimmerman, John and Ogan, Amy},
  title     = {"An Instructor is [already] able to keep track of 30 students": Students’ Perceptions of Smart Classrooms for Improving Teaching \& Their Emergent Understandings of Teaching and Learning},
  booktitle = {Proceedings of the 2023 ACM Designing Interactive Systems Conference},
  series    = {DIS '23},
  year      = {2023},
  isbn      = {9781450398930},
  location  = {Pittsburgh, PA, USA},
  pages     = {1277--1292},
  numpages  = {16},
  url       = {https://doi.org/10.1145/3563657.3596079},
  doi       = {10.1145/3563657.3596079},
  acmid     = {3596079},
  publisher = {ACM},
  address   = {New York, NY, USA}
}

@inproceedings{10.1145/3173574.3174091,
  author    = {Lechelt, Zuzanna and Rogers, Yvonne and Yuill, Nicola and Nagl, Lena and Ragone, Grazia and Marquardt, Nicolai},
  title     = {Inclusive Computing in Special Needs Classrooms: Designing for All},
  booktitle = {Proceedings of the 2018 CHI Conference on Human Factors in Computing Systems},
  series    = {CHI '18},
  year      = {2018},
  isbn      = {9781450356206},
  location  = {Montreal QC, Canada},
  pages     = {1--12},
  numpages  = {12},
  url       = {https://doi.org/10.1145/3173574.3174091},
  doi       = {10.1145/3173574.3174091},
  acmid     = {3174091},
  publisher = {ACM},
  address   = {New York, NY, USA}
}

@inproceedings{wang2020crescendo,
  author    = {Wengran Wang and Rui Zhi and Alexandra Milliken and Nicholas Lytle and Thomas W. Price},
  title     = {Crescendo: Engaging Students to Self‑Paced Programming Practices},
  booktitle = {Proceedings of the 51st ACM Technical Symposium on Computer Science Education (SIGCSE ’20)},
  year      = {2020},
  publisher = {ACM},
  doi       = {10.1145/3328778.3366919}
}

@inproceedings{10.1145/3173574.3173643,
  author    = {Milne, Lauren R. and Ladner, Richard E.},
  title     = {{Blocks4All}: Overcoming Accessibility Barriers to Blocks Programming for Children with Visual Impairments},
  booktitle = {Proceedings of the 2018 CHI Conference on Human Factors in Computing Systems},
  series    = {CHI '18},
  year      = {2018},
  location  = {Montreal, QC, Canada},
  articleno = {69},
  numpages  = {10},
  url       = {https://doi.org/10.1145/3173574.3173643},
  doi       = {10.1145/3173574.3173643},
  publisher = {ACM},
  address   = {New York, NY, USA},
  isbn      = {9781450356206}
}

@inproceedings{10.1145/3025453.3025711,
  author    = {Arawjo, Ian and Wang, Cheng-Yao and Myers, Andrew C. and Andersen, Erik and Guimbreti{\`e}re, Fran{\c{c}}ois},
  title     = {Teaching Programming with Gamified Semantics},
  booktitle = {Proceedings of the 2017 CHI Conference on Human Factors in Computing Systems},
  series    = {CHI '17},
  year      = {2017},
  location  = {Denver, Colorado, USA},
  pages     = {4911--4923},
  numpages  = {13},
  url       = {https://doi.org/10.1145/3025453.3025711},
  doi       = {10.1145/3025453.3025711},
  publisher = {ACM},
  address   = {New York, NY, USA},
  isbn      = {9781450346559}
}

@inproceedings{10.1145/3544549.3573863,
  author    = {Gennari, Rosella and Melonio, Alessandra and Rizvi, Mehdi},
  title     = {A Tool for Guiding Teachers and their Learners: the Case Study of an Art Class},
  booktitle = {Extended Abstracts of the 2023 CHI Conference on Human Factors in Computing Systems},
  series    = {CHI EA '23},
  year      = {2023},
  location  = {Hamburg, Germany},
  articleno = {376},
  numpages  = {6},
  url       = {https://doi.org/10.1145/3544549.3573863},
  doi       = {10.1145/3544549.3573863},
  publisher = {ACM},
  address   = {New York, NY, USA},
  isbn      = {9781450394222}
}

@inproceedings{10.1145/3544549.3585775,
  author    = {Limke, Ally and Lytle, Nicholas and Lin, Maggie and Mahmoud, Sana and Hill, Marnie and Catet{\'e}, Veronica and Barnes, Tiffany},
  title     = {Empowering Students as Leaders of Co-Design for Block-Based Programming},
  booktitle = {Extended Abstracts of the 2023 CHI Conference on Human Factors in Computing Systems},
  series    = {CHI EA '23},
  year      = {2023},
  location  = {Hamburg, Germany},
  articleno = {98},
  numpages  = {7},
  url       = {https://doi.org/10.1145/3544549.3585775},
  doi       = {10.1145/3544549.3585775},
  publisher = {ACM},
  address   = {New York, NY, USA},
  isbn      = {9781450394222}
}

@inproceedings{10.1145/3706599.3719763,
  author    = {Limke, Ally and Islam, Saminur and Riahi, Bahare and Tian, Xiaoyi and Hill, Marnie and Catet{\'e}, Veronica and Barnes, Tiffany},
  title     = {What Does It Take to Support Problem Solving in Programming Classrooms? A New Framework from the {{K-12}} Teacher Perspective},
  booktitle = {Proceedings of the Extended Abstracts of the CHI Conference on Human Factors in Computing Systems},
  series    = {CHI EA '25},
  year      = {2025},
  location  = {Yokohama, Japan},
  articleno = {591},
  numpages  = {7},
  url       = {https://doi.org/10.1145/3706599.3719763},
  doi       = {10.1145/3706599.3719763},
  publisher = {ACM},
  address   = {New York, NY, USA},
  isbn      = {9798400713958}
}

@inproceedings{10.1145/3643834.3661596,
  author    = {Zavaleta Bernuy, Angela and Sibia, Naaz and Chen, Pan and Xu, Jessica Jia-Ni and Tran, Elexandra and Ye, Runlong and Pammer-Schindler, Viktoria and Petersen, Andrew and Williams, Joseph Jay and Liut, Michael},
  title     = {Does the Medium Matter? An Exploration of Voice-Interaction for Self-Explanations},
  year      = {2024},
  isbn      = {9798400705830},
  publisher = {Association for Computing Machinery},
  address   = {New York, NY, USA},
  url       = {https://doi.org/10.1145/3643834.3661596},
  doi       = {10.1145/3643834.3661596},
  booktitle = {Proceedings of the 2024 ACM Designing Interactive Systems Conference},
  pages     = {86--101},
  numpages  = {16},
  location  = {Copenhagen, Denmark},
  series    = {DIS '24},
}

@inproceedings{10.1145/3613905.3650937,
  author    = {Xiao, Ruiwei and Hou, Xinying and Stamper, John},
  title     = {Exploring How Multiple Levels of {GPT}-Generated Programming Hints Support or Disappoint Novices},
  year      = {2024},
  isbn      = {9798400703317},
  publisher = {Association for Computing Machinery},
  address   = {New York, NY, USA},
  url       = {https://doi.org/10.1145/3613905.3650937},
  doi       = {10.1145/3613905.3650937},
  booktitle = {Extended Abstracts of the CHI Conference on Human Factors in Computing Systems},
  articleno = {142},
  numpages  = {10},
  location  = {Honolulu, HI, USA},
  series    = {CHI EA '24},
}

@inproceedings{10.1145/3411764.3445228,
  author    = {Weinman, Nathaniel and Fox, Armando and Hearst, Marti A.},
  title     = {Improving Instruction of Programming Patterns with Faded {Parsons} Problems},
  year      = {2021},
  isbn      = {9781450380966},
  publisher = {Association for Computing Machinery},
  address   = {New York, NY, USA},
  url       = {https://doi.org/10.1145/3411764.3445228},
  doi       = {10.1145/3411764.3445228},
  booktitle = {Proceedings of the 2021 CHI Conference on Human Factors in Computing Systems},
  articleno = {53},
  numpages  = {4},
  location  = {Yokohama, Japan},
  series    = {CHI '21},
}

@inproceedings{10.1145/2702123.2702580,
  author    = {O'Rourke, Eleanor and Andersen, Erik and Gulwani, Sumit and Popovi\'{c}, Zoran},
  title     = {A Framework for Automatically Generating Interactive Instructional Scaffolding},
  year      = {2015},
  isbn      = {9781450331456},
  publisher = {Association for Computing Machinery},
  address   = {New York, NY, USA},
  url       = {https://doi.org/10.1145/2702123.2702580},
  doi       = {10.1145/2702123.2702580},
  booktitle = {Proceedings of the 33rd Annual ACM Conference on Human Factors in Computing Systems},
  pages     = {1545--1554},
  numpages  = {10},
  location  = {Seoul, Republic of Korea},
  series    = {CHI '15},
}

@inproceedings{10.1145/3706598.3714002,
  author    = {Chen, Valerie and Zhu, Alan and Zhao, Sebastian and Mozannar, Hussein and Sontag, David and Talwalkar, Ameet},
  title     = {Need Help? Designing Proactive {AI} Assistants for Programming},
  year      = {2025},
  isbn      = {9798400713941},
  publisher = {Association for Computing Machinery},
  address   = {New York, NY, USA},
  url       = {https://doi.org/10.1145/3706598.3714002},
  doi       = {10.1145/3706598.3714002},
  booktitle = {Proceedings of the 2025 CHI Conference on Human Factors in Computing Systems},
  articleno = {881},
  numpages  = {18},
  series    = {CHI '25},
}

@inproceedings{10.1145/3313831.3376494,
  author    = {Lerner, Sorin},
  title     = {{Projection Boxes}: On-the-fly Reconfigurable Visualization for Live Programming},
  year      = {2020},
  isbn      = {9781450367080},
  publisher = {Association for Computing Machinery},
  address   = {New York, NY, USA},
  url       = {https://doi.org/10.1145/3313831.3376494},
  doi       = {10.1145/3313831.3376494},
  booktitle = {Proceedings of the 2020 CHI Conference on Human Factors in Computing Systems},
  pages     = {1--7},
  numpages  = {7},
  location  = {Honolulu, HI, USA},
  series    = {CHI '20},
}

@inproceedings{10.1145/3313831.3376857,
  author    = {Price, Thomas W. and Williams, Joseph Jay and Solyst, Jaemarie and Marwan, Samiha},
  title     = {Engaging Students with Instructor Solutions in Online Programming Homework},
  year      = {2020},
  isbn      = {9781450367080},
  publisher = {Association for Computing Machinery},
  address   = {New York, NY, USA},
  url       = {https://doi.org/10.1145/3313831.3376857},
  doi       = {10.1145/3313831.3376857},
  booktitle = {Proceedings of the 2020 CHI Conference on Human Factors in Computing Systems},
  pages     = {1--7},
  numpages  = {7},
  location  = {Honolulu, HI, USA},
  series    = {CHI '20},
}

@inproceedings{chatain2023embodied,
  author    = {Chatain, Julia and Kapur, Manu and Sumner, Robert W.},
  title     = {Three Perspectives on Embodied Learning in Virtual Reality: Opportunities for Interaction Design},
  booktitle = {Extended Abstracts of the 2023 CHI Conference on Human Factors in Computing Systems},
  series    = {CHI EA '23},
  year      = {2023},
  articleno = {281},
  numpages  = {8},
  url       = {https://doi.org/10.1145/3544549.3585805},
  doi       = {10.1145/3544549.3585805},
  publisher = {ACM},
  address   = {New York, NY, USA},
}

@inproceedings{revelle2005tui,
  author    = {Revelle, Glenda and Zuckerman, Oren and Druin, Allison and Bolas, Mark T.},
  title     = {Tangible user interfaces for children},
  booktitle = {Extended Abstracts of the 2005 CHI Conference on Human Factors in Computing Systems},
  series    = {CHI EA '05},
  year      = {2005},
  pages     = {2051--2052},
  numpages  = {2},
  url       = {https://doi.org/10.1145/1056808.1057095},
  doi       = {10.1145/1056808.1057095},
  publisher = {ACM},
  address   = {New York, NY, USA},
}

@inproceedings{zaidi2025tangibuild,
  author    = {Zaidi, Arooj and Barbareschi, Giulia and Sato, Chihiro and Yamaoka, Junichi},
  title     = {{TangiBuild}: A Tangible Learning Tool for Children's Structural Exploration with Real-Time Feedback},
  booktitle = {Extended Abstracts of the 2025 CHI Conference on Human Factors in Computing Systems},
  series    = {CHI EA '25},
  year      = {2025},
  articleno = {505},
  numpages  = {7},
  url       = {https://doi.org/10.1145/3706599.3720144},
  doi       = {10.1145/3706599.3720144},
  publisher = {ACM},
  address   = {New York, NY, USA},
}

@inproceedings{horn2006tangible,
  author    = {Horn, Michael S. and Jacob, Robert J. K.},
  title     = {Tangible programming in the classroom: a practical approach},
  booktitle = {Extended Abstracts of the 2006 CHI Conference on Human Factors in Computing Systems},
  series    = {CHI EA '06},
  year      = {2006},
  pages     = {869--874},
  numpages  = {6},
  url       = {https://doi.org/10.1145/1125451.1125621},
  doi       = {10.1145/1125451.1125621},
  publisher = {ACM},
  address   = {New York, NY, USA},
}

@inproceedings{horn2007tern,
  author    = {Horn, Michael S. and Jacob, Robert J. K.},
  title     = {Tangible programming in the classroom with tern},
  booktitle = {Extended Abstracts of the 2007 CHI Conference on Human Factors in Computing Systems},
  series    = {CHI EA '07},
  year      = {2007},
  pages     = {1965--1970},
  numpages  = {6},
  url       = {https://doi.org/10.1145/1240866.1240933},
  doi       = {10.1145/1240866.1240933},
  publisher = {ACM},
  address   = {New York, NY, USA},
}

@article{FeredayMuirCochrane2006,
  author = {Fereday, Jennifer and Muir-Cochrane, Eimear},
  title = {Demonstrating Rigor Using Thematic Analysis: A Hybrid Approach of Inductive and Deductive Coding and Theme Development},
  journal = {International Journal of Qualitative Methods},
  volume = {5},
  number = {1},
  pages = {80--92},
  year = {2006},
  doi = {10.1177/160940690600500107}
}

@article{chevalier2020fostering,
  title={Fostering computational thinking through educational robotics: A model for creative computational problem solving},
  author={Chevalier, Morgane and Giang, Christian and Piatti, Alberto and Mondada, Francesco},
  journal={International journal of STEM education},
  volume={7},
  number={1},
  pages={39},
  year={2020},
  publisher={Springer}
}

@article{chevalier2022role,
  title={The role of feedback and guidance as intervention methods to foster computational thinking in educational robotics learning activities for primary school},
  author={Chevalier, Morgane and Giang, Christian and El-Hamamsy, Laila and Bonnet, Evgeniia and Papaspyros, Vaios and Pellet, Jean-Philippe and Audrin, Catherine and Romero, Margarida and Baumberger, Bernard and Mondada, Francesco},
  journal={Computers \& Education},
  volume={180},
  pages={104431},
  year={2022},
  publisher={Elsevier}
}

@inproceedings{chevalier2021teachers,
  title={Teachers’ perspective on fostering computational thinking through educational robotics},
  author={Chevalier, Morgane and El-Hamamsy, Laila and Giang, Christian and Bruno, Barbara and Mondada, Francesco},
  booktitle={International Conference on Robotics in Education (RiE)},
  pages={177--185},
  year={2021},
  organization={Springer}
}

@article{weintrop2017comparing,
  title={Comparing block-based and text-based programming in high school computer science classrooms},
  author={Weintrop, David and Wilensky, Uri},
  journal={ACM Transactions on Computing Education (TOCE)},
  volume={18},
  number={1},
  pages={1--25},
  year={2017},
  publisher={ACM New York, NY, USA}
}

@inproceedings{kazemitabaar2022codestruct,
  title={Codestruct: Design and evaluation of an intermediary programming environment for novices to transition from scratch to python},
  author={Kazemitabaar, Majeed and Chyhir, Viktar and Weintrop, David and Grossman, Tovi},
  booktitle={Proceedings of the 21st Annual ACM Interaction Design and Children Conference},
  pages={261--273},
  year={2022}
}

@article{montuori2024cognitive,
  title={The cognitive effects of computational thinking: A systematic review and meta-analytic study},
  author={Montuori, Chiara and Gambarota, Filippo and Alto{\'e}, Gianmarco and Arf{\'e}, Barbara},
  journal={Computers \& Education},
  volume={210},
  pages={104961},
  year={2024},
  publisher={Elsevier}
}

@inproceedings{Wu2025TraceMate,
  author    = {Jinmiao Wu and Thomas Selig and Erick Purwanto},
  title     = {TraceMate: Collaborating with {AI} in Test-Driven Programming},
  booktitle = {Proceedings of the 2025 IEEE Symposium on Visual Languages and Human-Centric Computing (VL/HCC)},
  year      = {2025},
  pages     = {253--259},
  publisher = {IEEE Computer Society},
  doi       = {10.1109/VL-HCC65237.2025.00035}
}

@article{groothuijsen2024ai,
  title={AI chatbots in programming education: Students’ use in a scientific computing course and consequences for learning},
  author={Groothuijsen, Suzanne and Van den Beemt, Antoine and Remmers, Joris C and van Meeuwen, Ludo W},
  journal={Computers and Education: Artificial Intelligence},
  volume={7},
  pages={100290},
  year={2024},
  publisher={Elsevier}
}

@article{yang2024enhancing,
  title={Enhancing python learning with PyTutor: Efficacy of a ChatGPT-Based intelligent tutoring system in programming education},
  author={Yang, Albert CM and Lin, Ji-Yang and Lin, Cheng-Yan and Ogata, Hiroaki},
  journal={Computers and Education: Artificial Intelligence},
  volume={7},
  pages={100309},
  year={2024},
  publisher={Elsevier}
}

\clearpage


\appendix

\section{Prompt Used in the Call to the OpenAI API}
\label{app:prompt}

This section documents prompt structures and constraints used in calls to the OpenAI API (gpt-4o) for task generation and submission checking in our system.

\subsection{Tile Cleaning (Beginner)}

\begin{lstlisting}[style=promptblock]
You are an encouraging AI assistant that returns structured JSON outputs
for a Blockly-based learning system. 

Known blocks: ${block_info_json}

Task:
- Clean a 5×5 tiled floor area. Each tile has a side length of 300 mm.
- The robot starts from (0, 0), facing right (i.e., the +x direction).
  Coordinates are defined as (column, row).
- Package the action of cleaning two consecutive rows into a custom
  function named clean_2_rows (students may use a different function name).
- The body of clean_2_rows should consist of four steps:
  1) Move forward 4 tiles to clean one row;
  2) Move left to shift to the next row;
  3) Move backward 4 tiles to clean the next row;
  4) Move left again to prepare for the next iteration.
- In the main program, repeatedly call clean_2_rows twice using a loop,
  then move forward 4 tiles so that the robot covers the entire 5×5 grid.
- If students ask questions they do not understand, ask for answers,
  or show a negative attitude, respond positively with hints.
  Keep responses concise.
- If students ask whether their idea is correct and it is correct,
  provide brief confirmation.

Simulation Rules:
- <move_forward SPEED=X DURATION=T>:
  Move forward round(T × X / 300) tiles in the current direction.
- <move_left> / <move_right>:
  Translate laterally round(T × X / 300) tiles.
- <turn_left> / <turn_right>:
  Rotate in place.

Output Instructions:
1. When generating the problem (only once):
   - After a short opening sentence, directly output a detailed problem
     description, formatted as the following example JSON.

Example Output:
{
  "text": "problem description",
  "xml": "<xml xmlns=\"https://developers.google.com/blockly/xml\">...</xml>",
  "is_correct": false,
  "return_button": false
}

2. When analyzing a student's submitted Blockly XML:
- Parse the <xml> and simulate execution step by step by following the
  <next> chains and DO branches, tracking the complete (x, y, orientation)
  trajectory.
- Check:
  * Whether clean_2_rows is correctly defined;
  * Whether the correct loop count is selected.
- If correct:
  * Report the analysis result, congratulate the student, and end the
    exercise.
  * Set is_correct = true, return_button = true.
- Otherwise:
  * Follow the student's own reasoning and concisely explain the error.
  * Set is_correct = false, return_button = false.

Important:
- Strictly output JSON only. Do not include any additional text.
- Always compute movement distances using rounded values derived from
  SPEED and DURATION.
- Ensure the robot does not move outside the grid boundaries.
\end{lstlisting}

\subsection{Adventure in the Secret Realm (Intermediate)}

\begin{lstlisting}[style=promptblock]
You are an encouraging AI assistant that returns structured JSON outputs
for a Blockly-based learning system.

Known blocks: ${block_info_json}

Task Description:
- The robot starts from a fixed starting point S and must activate three
  triggers in the order A → B → C to open a sealed chamber.
- The environment is a 5×5 grid map, where each grid cell has a side length
  of 300 mm.
- The starting position S is fixed at (1,1), with the robot initially
  facing upward (positive y-axis direction).
- The trigger coordinates are:
  A = (4,3), B = (4,1), C = (4,2).
- Obstacle cells are located at:
  (0,4), (0,2), (1,3), (2,1), (3,1), and (3,3).
- A Blockly program is generated with a path that intentionally takes a
  detour. The example XML below activates the triggers in the correct
  order (A → B → C) but does not follow the shortest possible path.
  Students are asked to identify what is wrong with this solution.
- If a student points out an issue unrelated to path optimality, respond
  by clarifying that the core problem is the failure to find the shortest
  path.
- If students ask about unclear concepts, request the full solution, or
  show a negative attitude, respond positively and provide hints.
  Do not directly give the complete correct solution.

Movement and Boundary Rules (must be strictly followed):
- The grid size is 5×5, with coordinates x, y ∈ {0,1,2,3,4}.
  The origin (0,0) is the bottom-left corner. The x-axis increases to the
  right, and the y-axis increases upward.
- The side length of each grid cell is CELL_SIZE = 300 mm.
- For any movement block <move_* SPEED=X DURATION=T>, the number of grid
  cells moved is computed as:
  round((T × X) / 300).
- move_forward, move_backward, move_left, and move_right are translations
  relative to the robot’s current orientation and do not change its facing
  direction.
  * If orientation = North (↑):
    forward = (0,+n), backward = (0,-n),
    left = (-n,0), right = (+n,0)
  * If orientation = East (→):
    forward = (+n,0), backward = (-n,0),
    left = (0,+n), right = (0,-n)
  * West and South follow the same relative logic.
- After each movement step, the robot must remain within the bounds
  0 ≤ x ≤ 4 and 0 ≤ y ≤ 4. Any violation is treated as an out-of-bounds error.
- If a solution specifies only key waypoints (e.g., (1,2) → (4,2)), the
  system must expand the path into cell-by-cell movements and verify that
  no intermediate cell is blocked by an obstacle or exceeds the boundary.

Output Instructions:
1. When generating a new task:
   - After a short opening sentence, directly generate the full problem
     description, formatted as the following example.
   - Output ONE valid JSON object ONLY.

Example Output:
{
  "text": "problem description",
  "xml": "<xml xmlns=\"https://developers.google.com/blockly/xml\">...</xml>",
  "is_correct": false,
  "return_button": false
}

2. When analyzing a student's submitted Blockly XML:
   Task Evaluation Rules:
   - The correct path is defined as:
     (1,1) → (1,2) → (4,2) → (4,3) → (4,1),
     i.e., move up by one cell, move right by three cells, move up by one
     cell, move down by two cells, and then move up by one cell.
     (Turning instead of lateral translation is also allowed, as long as
     the same path is followed.)
   - If the student's path matches the correct path:
     * Report the analysis result and congratulate the student.
     * End the task and return to the interface.
     * Output:
       {
         "text": "feedback",
         "is_correct": true,
         "return_button": true
       }
   - Otherwise:
     * Set is_correct = false and return_button = false.
     * Provide a concise explanation of the error, following the student’s
       own reasoning.

\end{lstlisting}

\subsection{Mineral Collection (Intermediate)}

\begin{lstlisting}[style=promptblock]
You are an encouraging AI assistant that returns structured JSON outputs
for a Blockly-based learning system.

Known blocks: ${block_info_json}

Task:
- The environment is a 5×5 grid map. Each grid cell has a side length of
  300 mm. The bottom-left corner of the map is defined as (0,0), with the
  x-axis increasing to the right and the y-axis increasing upward.
- The robot starts at position (0,0) with an initial orientation facing
  upward (i.e., +y direction). Coordinates are defined as (column, row).
- Mineral locations are at: (1,0), (1,3), and (3,3).
- Swamp locations are at: (0,2), (2,2), (2,3), and (2,4).
- The robot starts with an initial energy level of 6.
  * Moving one grid cell consumes 1 unit of energy.
  * Collecting a mineral consumes 1 unit of energy.
  * Successfully collecting a mineral restores 3 units of energy.
  * Entering a swamp consumes an additional 2 units of energy.
- The robot must collect all minerals while its energy level remains
  greater than 0.
- Energy changes are computed automatically when the robot passes through
  a cell:
  * Passing through a mineral cell triggers automatic collection.
  * Passing through a swamp cell triggers an energy penalty.
- An incorrect route is intentionally designed as follows:
  move right by 3 cells, move forward by 3 cells, then move left by 3 cells.
  Under this route, the robot reaches the third mineral with zero remaining
  energy and thus cannot collect it. Students are required to identify
  the mistake and propose a correct route.
- If students ask about unclear concepts, request the answer directly,
  or show a negative attitude, respond positively with hints.
  Do not directly provide the full correct solution.

Simulation Rules:
- move_forward, move_backward, move_left, and move_right are all defined
  as translations relative to the robot's current orientation and do not
  change its facing direction.
- <move_forward SPEED=X DURATION=T>:
  Move forward round(T × X / 300) grid cells.
- <move_backward SPEED=X DURATION=T>:
  Move backward round(T × X / 300) grid cells.
- <turn_left / turn_right DEGREES=90>:
  Change the robot's orientation.

Output Instructions:
1. When generating a new task:
   - After a short opening sentence, directly generate the full problem
     description, formatted as the following example.
   - Output ONE valid JSON object ONLY.

Example Output:
{
  "text": "problem description",
  "xml": "<xml xmlns=\"https://developers.google.com/blockly/xml\">...</xml>",
  "is_correct": false,
  "return_button": false
}

2. When analyzing a student's submitted Blockly XML:
- Strictly parse the <xml> and recursively execute blocks following the
  <next> chain.
- For each execution step:
  * move_forward / move_backward / move_left / move_right:
    update the robot position according to the formula.
  * turn_left / turn_right:
    update the robot orientation.
- Strictly track (x, y, orientation).
- After executing all blocks:
  * If the solution is correct, report the analysis result, congratulate
    the student, and end the exercise.
    Set is_correct = true, return_button = true.
  * If the solution is incorrect, set is_correct = false and
    return_button = false, and concisely explain the error based on the XML.

Important:
- Always compute movement distances strictly using SPEED and DURATION.
- Ensure that the robot never moves outside the grid boundaries.
- Output JSON ONLY. Do not include any explanation outside the JSON object.
\end{lstlisting}

\subsection{Wolf, Goat, and Cabbage (Advanced)}

\begin{lstlisting}[style=promptblock]
You are an encouraging AI assistant that returns structured JSON outputs
for a Blockly-based learning system. 

Known blocks: ${block_info_json}

Task Description:
- The environment is a 5×5 grid map, where each grid cell has a side length
  of 300 mm.
- The robot initially starts at position (0,0), which is the bottom-left
  corner of the map, and faces left (i.e., the negative x-axis direction).
- Three items are placed outside the grid on the left side. When the robot
  is located at (0,0) and facing left, it can directly pick up one of these
  items, carry it into the grid, and later place it down.
- The robot can carry only one item at a time.
- The river spans three consecutive grid cells: (0,0), (1,0), and (2,0).
- When the robot is located at (2,0), it can use a “place” action to place
  the carried item onto the opposite bank at position (3,0).
- The opposite bank position (3,0) is a valid placement location for items,
  but the robot itself is not allowed to enter (3,0).
- Therefore, the placement action must occur while the robot is at (2,0),
  and the placed item must reach (3,0).
- During the river-crossing process, the robot may either:
  * rotate 180 degrees and then move forward, or
  * move backward first and then rotate 180 degrees.
- If students ask about unclear concepts, request the answer directly, or
  show a negative attitude, respond positively and provide hints.

Simulation Rules:
- move_forward, move_backward, move_left, and move_right are translations
  relative to the robot’s current orientation and do not change its facing
  direction.
- <move_forward SPEED=X DURATION=T>:
  Move forward round(T × X / 300) grid cells in the current orientation.
- <move_backward SPEED=X DURATION=T>:
  Move backward round(T × X / 300) grid cells relative to the current
  orientation.
- <turn_left / turn_right DEGREES=90>:
  Change the robot’s orientation.
- Ensure that the robot never moves outside the grid boundaries.

Output Instructions:
1. When generating a new task:
   - Output ONE valid JSON object ONLY.
   - After a short opening sentence, directly generate the full problem
     description, formatted as the following example.

Example JSON Output:
{
  "text": "problem description",
  "xml": null,
  "is_correct": false,
  "return_button": false
}

2. When analyzing a student's submitted Blockly XML:
- Strictly parse the <xml> and recursively execute blocks following the
  <next> chain.
- For each execution step:
  * move_forward / move_backward / move_left / move_right:
    update the robot’s position according to the movement formula.
  * turn_left / turn_right:
    update the robot’s orientation.
- Strictly track (x, y, orientation).
- Analyze the entire XML before making a final judgment:
  * If the solution is correct:
    - Congratulate the student and end the exercise.
    - Set is_correct = true and return_button = true.
  * If the solution is incorrect:
    - Set is_correct = false and return_button = false.
    - Provide a concise explanation of the error based on the XML.

Important:
- Always compute movement distances strictly using SPEED and DURATION.
- Ensure that the robot never moves outside the grid boundaries.
- Output JSON ONLY. Do not include any explanation outside the JSON object.
\end{lstlisting}

\subsection{Knight's Tour (Advanced)}
\begin{lstlisting}[style=promptblock]
You are an encouraging AI assistant that returns structured JSON outputs
for a Blockly-based learning system.

Known blocks: ${block_info_json}

Task Description:
- The robot starts from a fixed starting point S and must move using only
  knight-style (L-shaped) moves to cover all grid cells except the start
  cell and obstacle cells, and finally reach a specified goal cell.
  Each grid cell has a side length of 300 mm.
- The starting position S is at (0,0), with the robot initially facing
  upward (positive y-axis direction). Coordinates are defined as
  (column, row).
- The goal position G is at (3,0).
- Obstacle cells are located at (3,1), (4,3), and (4,4).
- Each move must follow a knight pattern:
  either two cells horizontally and one cell vertically, or two cells
  vertically and one cell horizontally.
- Except for obstacle cells, every grid cell must be visited exactly once.
  The starting cell is considered already covered and must not be revisited.
- One valid reference path is:
  (0,0),
  {(2,0),(2,1)},
  {(0,1),(0,2)},
  {(0,4),(1,4)},
  {(1,2),(2,2)},
  {(2,4),(3,4)},
  {(3,2),(4,2)},
  {(4,0),(3,0)}.
- If students ask for the correct answer, do not directly provide the full
  path. You may hint at one or two steps or suggest using translation blocks
  to simplify movement.
- If students appear confused or show a negative attitude, respond with
  encouragement and hints, but do not directly provide the complete solution.

Simulation Rules:
- move_forward, move_backward, move_left, and move_right are translations
  relative to the robot’s current orientation and do not change its facing
  direction.
- move_knight:
  * First, check legality:
    - The move must consist of two cells in one axis and one cell in the
      perpendicular axis.
    - Movement is defined relative to the robot’s current orientation.
  * Convert relative directions into a global displacement vector (Δx, Δy).
  * Merge horizontal and vertical components before legality checking.
    A move is legal if and only if (|Δx|, |Δy|) is (1,2) or (2,1).
  * Expand the knight move into two basic translation steps (e.g., horizontal
    first, then vertical). Each intermediate cell traversed must be recorded.
  * If the move is illegal, record the error reason but continue the full
    analysis.
- turn_left / turn_right:
  Change the robot’s orientation.
- At each step, strictly track (x, y, orientation) and ensure the robot
  does not move outside the grid boundaries.

Output Instructions:
1. When generating a new task:
   - Output ONE valid JSON object ONLY.
   - After a short opening sentence, directly generate the following example
     problem text (the task description is generated only once).

Example JSON Output:
{
  "text": "problem description",
  "xml": null,
  "is_correct": false,
  "return_button": false
}

2. When analyzing a student's submitted Blockly XML:
- Fully parse the <xml>, recursively processing <next> chains and
  <statement name="DO"> blocks.
- move_forward / move_backward / move_left / move_right:
  move round(DURATION × SPEED / 300) grid cells.
- For move_knight:
  * DIR_X (left/right) and DIR_Y (forward/backward) are relative to the
    robot’s current orientation.
  * Convert relative directions into global displacement vectors before
    computing Δx and Δy.
  * Direction mapping example when orientation is North (↑ / +y):
    forward = (0,+1), backward = (0,-1),
    right = (+1,0), left = (-1,0).
    For other orientations, apply the corresponding rotation.
  * For each move_knight instruction, first combine horizontal and vertical
    components into a single global displacement (Δx, Δy) and then check
    legality. Do not judge legality step-by-step.
  * When expanding a legal move, verify that each intermediate cell does
    not exceed boundaries, is not an obstacle, and has not been previously
    visited (if intermediate cells are counted as visited).
- turn_left / turn_right:
  update the robot’s orientation.
- Update (x, y, orientation) and record all visited cells.
- After completing the analysis, determine correctness:
  * If correct:
    - Congratulate the student.
    - Set is_correct = true and return_button = true.
  * If incorrect:
    - Set is_correct = false and return_button = false.
    - Provide concise feedback explaining the error.

Important:
- Always compute movement distances strictly using SPEED and DURATION.
- Ensure that the robot never moves outside the grid boundaries.
- Output JSON ONLY. Do not include any explanation outside the JSON object.
\end{lstlisting}

\section{Interview Protocol}
\label{app:interview}

We conducted a semi-structured teacher interview to understand classroom practices, constraints, and expectations for AI-supported, block-based robotics learning. The interview focused on how teachers support computational thinking (CT), make code–behavior relationships observable, and define appropriate boundaries for AI assistance in classroom settings.  
Table~\ref{tab:interview-questions} summarizes the interview themes and guiding prompts.

\begin{table}[t]
\centering
\caption{\rebrep{Overview of the semi-structured teacher interview protocol.}{Overview of the semi-structured teacher interview protocol and example prompts.}}
\label{tab:interview-questions}
\scriptsize
\setlength{\tabcolsep}{2pt}
\renewcommand{\arraystretch}{1.05}
\begin{tabular}{@{}p{0.32\columnwidth}p{0.62\columnwidth}@{}}
\textbf{Theme} & \textbf{\rebrep{Guiding prompts (summary)}{Example interview prompts}} \\
\midrule

Classroom context \& CT goals &
Teaching context, including age or grade level, class size, lesson length, and platform; up to three prioritized CT outcomes; observable evidence used to judge achievement, such as behaviors, artifacts, explanations, or assessments. \\

\midrule
Making program behavior visible &
A recent moment when running code clarified student understanding; available feedback or representations, such as physical execution, simulation, traces, or step-through views; what became inspectable when linking code and behavior. \\

\midrule
Embodied--digital mappings &
Key mappings between programming constructs and embodied actions, such as variables to parameters, conditionals to thresholds, and loops to repetition; which mappings are intuitive or misconception-prone, with examples. \\

\midrule
Learner control \& programmability &
What students should be able to modify, such as parameters, blocks, control flow, logic from scratch, or tests; which decisions should remain student-controlled rather than automated; how this supports reasoning rather than answer consumption. \\

\midrule
AI assistance and agency &
An ideal AI intervention sequence when students are stuck, such as diagnosing the problem, surfacing evidence, offering options, and providing solutions only as a last resort; examples of interventions that may reduce student agency. \\

\midrule
Explainability \& transparency &
What AI help should make transparent, including the reasoning behind hints, error causes, alternatives, verification, and uncertainty; outputs to avoid in classrooms, such as full solutions without rationale. \\

\midrule
Debugging difficulty \& iteration &
The most challenging debugging steps for students; useful slow-down supports, such as pause, step-through, replay, traces, and version comparison; a scenario where such support changes reasoning or interaction. \\

\midrule
Task progression &
An example programming task with a three-stage progression; learning goals at each stage; how complexity increases; strategies to prevent students from falling behind, such as scaffolds, checkpoints, or exemplars. \\

\midrule
Classroom orchestration &
Desired teacher-facing supports, such as real-time progress, common error hotspots, and help-seeking signals; how and when teachers would use them; practical classroom constraints. \\

\midrule
Ownership of code \& outcomes &
How students demonstrate ownership, such as explanations, version evolution, self-designed tests, or demonstrations; rules or mechanisms to preserve ownership when AI support is present. \\

\bottomrule
\end{tabular}
\end{table}

\section{Student Interview Questions}

To better understand how students interact with RoboBlockly Studio, we prepared questions covering six thematic areas and conducted brief interviews in which each student responded to two to three selected questions based on their completed experiences. Sample questions are provided in Table~\ref{tab:student-interviews}.

The contents of the NASA-TLX, SUS, UEQ, and Bebras questions were not modified for this study and are therefore not included in the Appendix.

\begin{table}[t]
\centering
\caption{Interview Themes and Sample Questions for Students}
\label{tab:student-interviews}
\scriptsize
\setlength{\tabcolsep}{2pt}
\renewcommand{\arraystretch}{1.08}
\begin{tabular}{@{}p{0.34\columnwidth}p{0.60\columnwidth}@{}}
\toprule
\textbf{Theme} & \textbf{Sample Question} \\
\midrule

Robot Embodiment &
When observing the robot executing a task, are you able to identify errors in your program? \\

\midrule
Debugging Strategies &
When encountering a program error, do you first observe the robot or ask the AI for help? Why? \\

\midrule
Verification and Error Correction &
After completing a task, how do you usually verify that your program is correct? \\

\midrule
AI Support and Iterative Development &
After observing the robot's execution, how did the questions you asked the AI change? \\

\midrule
Visual Programming / Block-based Interaction &
Do you find constructing programs by dragging blocks more intuitive than asking the AI using text? Why? \\

\midrule
Computational Thinking Skills &
In multi-step tasks, how do you break down problems into manageable sub-tasks? \\

\bottomrule
\end{tabular}
\end{table}


\end{document}